\documentclass[acmlarge, prologue, table]{acmart}

\usepackage{graphicx}
\usepackage{subcaption}
\usepackage[normalem]{ulem}
\usepackage[ruled]{algorithm2e} % For algorithms

\usepackage{makecell}
\usepackage{multicol}
\usepackage{multirow}
\usepackage{tcolorbox}

\AtBeginDocument{%
  \providecommand\BibTeX{{%
    \normalfont B\kern-0.5em{\scshape i\kern-0.25em b}\kern-0.8em\TeX}}}

\setcopyright{acmcopyright}
\copyrightyear{2024}
\acmYear{2024}
\acmDOI{XXXXXXX.XXXXXXX}

\begin{document}

\title{Thoughtful Things: Building Human-Centric Smart Devices with Small Language Models}


\author{Evan King}
\email{e.king@utexas.edu}
\orcid{0000-0003-4689-4591}
\affiliation{
  \institution{University of Texas at Austin}
  \city{Austin}
  \state{Texas}
  \country{USA}
}

\author{Haoxiang Yu}
\email{hxyu@utexas.edu}
\orcid{0000-0002-3518-946X}
\affiliation{
  \institution{University of Texas at Austin}
  \city{Austin}
  \state{Texas}
  \country{USA}
}

\author{Sahil Vartak}
\email{sv25972@utexas.edu}
% \orcid{0000-0002-3518-946X}
\affiliation{
  \institution{University of Texas at Austin}
  \city{Austin}
  \state{Texas}
  \country{USA}
}

\author{Jenna Jacob}
\email{jenna.jacob@utexas.edu}
% \orcid{0000-0002-3518-946X}
\affiliation{
  \institution{University of Texas at Austin}
  \city{Austin}
  \state{Texas}
  \country{USA}
}

\author{Sangsu Lee}
\email{sethlee@utexas.edu}
\orcid{0000-0002-3348-3163}
\affiliation{
  \institution{University of Texas at Austin}
  \city{Austin}
  \state{TX}
  \country{USA}
}

\author{Christine Julien}
\email{c.julien@utexas.edu}
\orcid{0000-0002-4131-4642}
\affiliation{
  \institution{University of Texas at Austin}
  \city{Austin}
  \state{Texas}
  \country{USA}
}

\renewcommand{\shortauthors}{E. King, et al.}

\begin{abstract}
Everyday devices like light bulbs and kitchen appliances are now embedded with so many features and automated behaviors that they have become complicated to actually \emph{use}. While such ``smart'' capabilities can better support users' goals, the task of learning the ``ins and outs'' of different devices is daunting. Voice assistants aim to solve this problem by providing a natural language interface to devices, yet such assistants cannot understand loosely-constrained commands, they lack the ability to reason about and explain devices' behaviors to users, and they rely on connectivity to intrusive cloud infrastructure. Toward addressing these issues, we propose \emph{thoughtful things}: devices that leverage lightweight, on-device language models to \emph{take actions} and \emph{explain their behaviors} in response to unconstrained user commands. We propose an end-to-end framework that leverages formal modeling, automated training data synthesis, and generative language models to create devices that are both capable and \emph{thoughtful} in the presence of unconstrained user goals and inquiries. Our framework requires no labeled data and can be deployed on-device, with no cloud dependency. We implement two thoughtful things (a lamp and a thermostat) and deploy them on real hardware, evaluating their practical performance.
\end{abstract}

\begin{CCSXML}
\end{CCSXML}


\keywords{natural language interfaces, generative language models, self-explaining systems, smart devices, human-centered computing}



\maketitle

\section{Introduction}
\label{sec:introduction}
Miniaturization of hardware, coupled with ubiquitous connectivity and increasingly powerful software implementations have produced a dramatic boost in the capability---and complexity---of everyday devices. Ovens can now subtly adapt cooking modes to different recipes~\cite{li2009intelligent}, while thermostats can utilize long-term sensing to learn energy-efficient schedules based on user habits~\cite{googlenest}. Even light bulbs have evolved from simple ``on-off'' devices to include the ability to adapt to a user's circadian rhythms~\cite{kwon2017multi}. This complexity is further compounded in high-end ``prosumer'' devices, which tend to incorporate even deeper sets of features---from musical synthesizers with nearly-infinite combinations of parameters~\cite{shier2021synthesizer} to exercise bikes that adapt workout routines to different fitness levels~\cite{bell_erg_nodate}. However, for all but the most technically literate ``power users'', these smart devices can be intimidating~\cite{iancu2020love, he2019smart}---if only because their capabilities are not easily understood, and the reasoning behind their behaviors is often unclear~\cite{yang2013learning}. In understanding the capabilities and behaviors of their devices, users face a steeper learning curve than ever: the user manual for a ``smart'' General Electric refrigerator, for instance, is twice as long as its ``non-smart'' counterpart~\cite{ge2023smart, ge2023nonsmart}. Modern devices are empowered to do more than ever and, as a result, are more complicated than ever to actually \emph{use}. 

Natural language interfaces (NLI) are a popular way for users to control smart devices. The most widely-adopted NLI are ``smart home assistants'' like Alexa or Google Home, which can interpret simple voice commands to control devices connected to a user's home network. In practice, home assistants have numerous issues; among them, they impose burdensome, rigid command structures that make them frustrating to use~\cite{upadhyay2023studying, luger2016like} and rely on constant connectivity to the cloud. With cloud dependency comes a host of other concerns: high business cost~\cite{ciuffoletti2018low}, weak user privacy~\cite{tawalbeh2020iot, barbosa2020privacy}, and unpredictable ``bricking'' of users' devices when services are sunset~\cite{parab_google_2024, camp_my_nodate, ashworth_smart_nodate}. While the recent introduction of large language models (LLMs) for device control provides added flexibility~\cite{king2023get}, these approaches rely on GPU-hungry LLMs that demand either powerful hardware~\cite{paul2024enabling} or impose the same cloud dependency as existing smart home assistants~\cite{king2024sasha}. Furthermore, recent LLM-based smart home systems cannot explain the reasoning behind devices' behaviors---they can act, but they cannot \emph{think}.

\begin{figure}[t!]
    \centering
    \includegraphics[width=\textwidth]{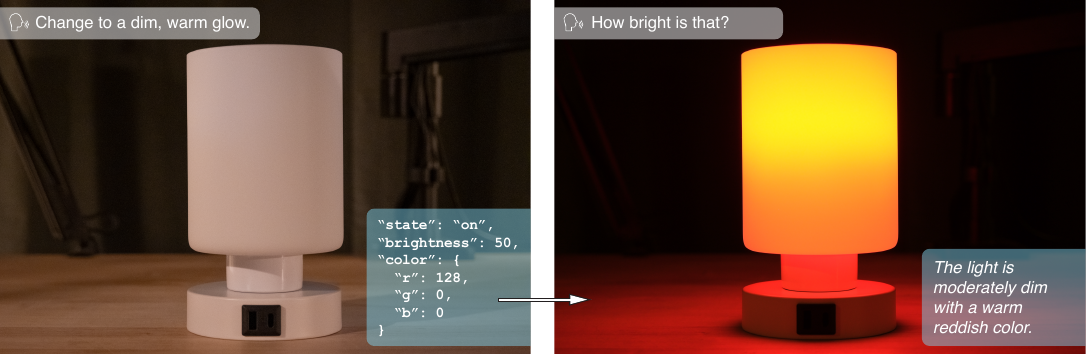}
    \caption{Our prototype implementation of a ``thoughtful lamp''. The lamp leverages a fine-tuned small language model running on a Raspberry Pi to respond to loosely constrained user commands with appropriate actions (e.g., changing color; left to right) and explanations (e.g., describing its state and capabilities; right)}
    \label{fig:teaser}
\end{figure}

In this paper, our objective is to build smart devices that are not only capable, but also \emph{thoughtful.} This ``thoughtfulness'' refers to individual devices that possess some degree of \emph{self-knowledge} that allows them to flexibly reason about their internal settings and states in relation to users' inquiries and goals. A user might ask a thoughtful lamp, for instance, to ``make it more welcoming in here'', and the lamp can reason about ``welcoming'' in relation to its available settings to meet the user's goal in a flexible manner. A thoughtful thermostat might leverage its self-knowledge to explain to a confused user who asks ``why did the heat turn off?'' when the thermostat executes an otherwise-opaque learned schedule. In realizing a \emph{thoughtful thing}, we synergize a feature-rich device's ability to take \emph{actions} and provide helpful \emph{explanations} in response to loosely-constrained user commands. We accomplish this with \emph{no dependence on the cloud or on GPU-rich local hardware}, which improves device longevity and benefits user privacy. Consequently, these devices are more human-centric because user interactions are tailored to the goals, communication styles, and privacy concerns of users.

% \slee{Do we need to provide a context of how a thoughtful thing would interoperate with multiple thoughtful things? If we have current implementations of thoughtful lamps and a thoughtful thermostat, they will try to answer the user's question regardless of whether it's about the lamp or thermostat. What does a smart home system of thoughtful things look like? If it's outside of the scope of this work, how can we improve this in the future?}

The key insight of our work is that we only need to define a \emph{state model} for a device, after which we can automate the task of training a small language model to reason about the device in the manner of a ``thoughtful thing''. We develop this concept in several steps. First, we establish a definition of ``thoughtful things'', along with some simplifying assumptions. We then introduce a general-purpose framework for building thoughtful things. Our five-step approach allows us to build thoughtful things by training a \emph{grounded, lightweight model of self-knowledge} using a combination of formal modeling, automated training data synthesis, and small (i.e., sub-3B-parameter), fine-tuned language models. This self-knowledge empowers a thoughtful thing to reason about its internal settings and states in relation to unconstrained natural language commands given by users without a need for cloud connectivity or expensive hardware, and requires \emph{no labeled data collection} to train. To demonstrate our framework, we undertake two case study implementations of thoughtful things---a lamp and a thermostat. We deploy both implementations on a realistic consumer device (Raspberry Pi) and show the effectiveness of our framework for building devices that provide high-quality, human-centric user interactions.

Our key contributions are as follows:

\begin{itemize}
    \item A practical framework for building \emph{thoughtful things}: devices that leverage small, fine-tuned language models to reason about their own behaviors and capabilities to respond to unstructured user commands with appropriate \emph{actions} or \emph{explanations}. Our framework \emph{requires no labeled training data} and leverages models small enough to run on-device, \emph{removing the need for costly, intrusive, and ephemeral cloud infrastructure.}
    \item A set of evaluations that demonstrate the practical performance of models produced by our framework.
    \item Two implementations of thoughtful things---a thoughtful lamp and a thoughtful thermostat---which we integrate on readily available consumer hardware (Raspberry Pi 5). We collect empirical measurements of on-device performance (e.g., response latency), which hint at feasibility for practical applications.
\end{itemize}

The paper is organized as follows. Section~\ref{sec:background} situates the paper with related work, definitions, and simplifying assumptions. Section~\ref{sec:framework} introduces our framework for building thoughtful things. Section~\ref{sec:implementations} describes our real-world implementations of two thoughtful things using our framework: a thoughtful lamp and thoughtful thermostat. Section~\ref{sec:evaluation} evaluates the quality and runtime performance of the models produced by our framework, while Section~\ref{sec:discussion} discusses limitations and avenues for future work. Section~\ref{sec:conclusion} concludes.

\section{Background}
\label{sec:background}
We first discuss related work in empirical studies, self-explaining systems, natural language interfaces, and large language models. We then establish a set of definitions and assumptions that situate the remainder of the paper.
\subsection{Related Work}
\textbf{Prior empirical studies} motivate our effort to improve smart device usability and understandability. Lazar et al. found that smart device users ``may not have the expertise of how devices could suit their goals'' and that ``extra work...was a significant issue'' that leads to abandonment of smart devices~\cite{lazar2015we}. He et al. noted that smart device users often believe ``one had to be an expert to write rules'' to make full use of one's devices, and that ``inexperienced users were intimidated''. They also found that the opaque automated behaviors of, e.g., smart thermostats, are a common cause of frustration, concluding that ``it is necessary to give users greater transparency'' about devices' behaviors~\cite{he2019smart}. The importance of transparency is echoed in studies by Coskun et al. and Yang et al.~\cite{coskun2018smart, yang2013learning}. In a study among older adults, Iancu and Iancu found that while users ``like owning smart devices, they consider them too complex and intrusive'', and concluded that users ``require substantial help in using, understanding and discovering new functions of...device[s]''~\cite{iancu2020love}. Privacy is particularly important, as user fears about cloud-connected devices that are ``always listening'' are known to reduce adoption~\cite{gudovskiy2019smart}. User studies signal a need for smart devices that are \emph{usable}, \emph{understandable}, and \emph{unintrusive} in spite of their complexity~\cite{li2021motivations}.

\textbf{Self-explaining systems} that can describe their capabilities and reasoning are proposed to improve user trust and understanding~\cite{dey2009explanations, kulesza2013too}. In smart environments, prior work achieves self-explanation by composing formal descriptions of device capabilities and/or automation rules to provide instructions and descriptions to users. Lieberman et al. proposed Roadie, a framework for generating device instructions by leveraging a knowledge graph of user goals, a set of device actions, and a partial-order planner~\cite{lieberman2006goal}. Burmeister et al. subsequently proposed a ``Smart Object Description Language'' for formalizing the capabilities of smart devices~\cite{burmeister2015ambient, burmeister2017smart}, which Kordts et al. integrated in a framework that synthesizes user documentation for ensembles of smart devices~\cite{kordts2021towards}. Coppers et al. and Sadeghi et al. similarly leverage formal treatments for self-explanation, but focus on user-centric explanations of automation routines at different levels of specificity~\cite{coppers2020fortniot, sadeghi2024smartex}. As in prior work, our framework leverages formal descriptions of devices---albeit to a different end. We leverage formalization as a sort of grounding for training and reasoning with a generative language model. This allows us to incorporate out-of-domain knowledge---e.g., information about the how device settings relate to abstract concepts like mood, or human contexts---that tends to be absent from formal specifications.

\textbf{Natural language interfaces (NLI)} that enable users to interact with devices via spoken commands are popular due to their purported accessibility and ease-of-use~\cite{pradhan2018accessibility, aung2018}. Yates et al. proposed an early NLI for appliances that leverages a behavioral model and relational database to allow users to query or modify device state with structured commands, e.g., to ask an answering machine ``how many messages?'' or to ``delete old messages''~\cite{yates2003reliable}. A variety of works have proposed NLI to other devices, e.g., fridges~\cite{ferrero2019ubiquitous, gudovskiy2019smart}, automobiles~\cite{tsiakoulis2012statistical}, and sound synthesizers~\cite{meinicke2020multi, brade2024synthscribe}. Research in NLI has predominantly focused on methods for \emph{intent detection} and \emph{slot filling}, through which a system extracts key intents and entities from user utterances to map to hard-coded system actions~\cite{weld2022survey}. ``Smart home assistants'' like Google Home and Alexa are arguably the dominant form of NLI for smart device control in practice. These assistants tend to suffer from an inability to understand loosely-constrained commands, which contributes to high user friction and dissatisfaction while limiting accessibility~\cite{luger2016like, cowan2017can, pradhan2020use}.

\begin{figure}
    \centering
    \begin{subfigure}[b]{0.25\textwidth}
        \centering
        \includegraphics[width=\textwidth]{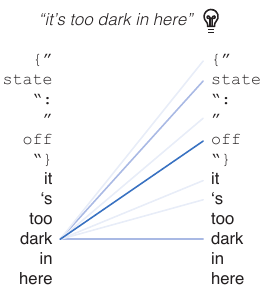}
    \end{subfigure}
    \hspace{1cm}
    \begin{subfigure}[b]{0.307\textwidth}
        \centering
        \includegraphics[width=\textwidth]{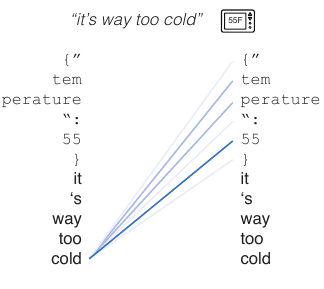}
    \end{subfigure}
    \caption{Visualizations of attention in a transformer model (Phi-2) given input user commands and device states. Generalist base models are pre-trained on large amounts of code and unconstrained text, so they learn semantic relationships between commands and relevant machine-readable state (e.g., ``dark'' and \texttt{``off''}). In this paper, we fine-tune these models to generate responses with device-specific actions and explanations that adhere to a real device's state model.}
    \label{fig:attention}
\end{figure}

\textbf{Large language models (LLMs)} have recently gained traction due to their high performance at many downstream natural language tasks. LLM research stems from the introduction of the transformer architecture by Vaswani et al., which leverages ``attention'' to model inter-sequence dependencies in language~\cite{vaswani2017attention}. When trained on significant amounts of unconstrained text, these models can generalize to a variety of tasks involving language, including code synthesis~\cite{xu2022systematic} and reasoning~\cite{kojima2022large}. These models also exhibit good performance at \emph{smart home control}, as they are able to model the relationship between unconstrained natural language and device settings and/or state encoded in, e.g., JSON (see Fig.~\ref{fig:attention}). King et al. introduced the use of LLMs for controlling ensembles of smart devices with Sasha, a system that leverages general-purpose LLMs (i.e., GPT-4~\cite{achiam2023gpt} and Llama 2~\cite{touvron2023llama}) to generate ``action plans'' in response to unconstrained user commands~\cite{king2024sasha}. Paul et al. proposed a method for synthesizing data and fine-tuning smaller language models (i.e., Llama 2) to specialize at smart device control~\cite{paul2024enabling}. As with prior work, we leverage language models for device control, but introduce several important novelties. First, our work leverages local models that are less than half the size of the state of the art for smart spaces---the 2.7B Phi-2~\cite{gunasekar2023textbooks} and the 2B Gemma~\cite{team2024gemma} base models---which greatly reduces the cost of training and implementation in real devices. Second, our approach differs in that we train models to reason about the state of individual devices, rather than ensembles. Finally, our work is the first to introduce the use of small language models for smart device \emph{explainability}, which has not been explored in the literature.

\subsection{Definitions \& Assumptions}
\label{sec:defintion}
\begin{figure}
    \centering
    \includegraphics[width=\textwidth]{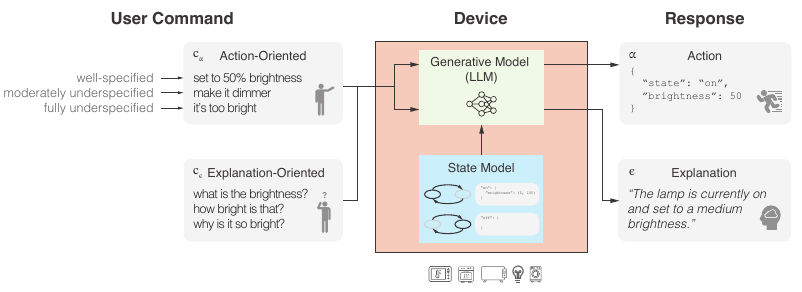}
    \caption{Thoughtful things are devices that respond to unconstrained user commands with appropriate actions (i.e., state changes) or explanations (i.e., descriptions of current state and capabilities). We accomplish this by combining a small, fine-tuned generative language model with a formal system model. The LLM flexibly synthesizes new states and explanations in response to diverse user commands, while the system model grounds responses in a device's true capabilities.}
    \label{fig:overview}
\end{figure}

Our goal is to build everyday devices that can respond to a given unconstrained user command with either an appropriate \textbf{action} $\alpha$ or logically correct \textbf{explanation} $\epsilon$ (Fig.~\ref{fig:overview}). A device \emph{acts} by transitioning to a valid and appropriate state, i.e., by choosing settings that satisfy the user's request. An action $\alpha$ is therefore an actionable (i.e., machine-parseable) description of the state that a device should take in response to a command. A device \emph{explains} by generating a natural language description of its current state, with sensitivity to the device's capabilities. Taking the example of a ``thoughtful lamp'', the device should respond to an \textbf{action-oriented command} $c_\alpha$ like ``make the light better for reading'' by generating an action $\alpha$ that assigns, e.g., a brightness and color that a user would find appropriate for reading. If the user issues an \textbf{explanation-oriented command} $c_\epsilon$ like ``is this as bright as it gets?'', the lamp should reason about its capabilities in relation to its current state to produce an explanation $\epsilon$, e.g., ``No, the light is at half brightness, so it can be brighter.''

Natural language commands have varying levels of \textbf{specificity}. We adapt a definition of command specificity from related work~\cite{king2024sasha}. We target three levels:

\begin{itemize}
    \item \textbf{Well-specified commands} describe specific, desired actions/explanations and relevant settings, e.g., ``set to 50\% brightness'' or ``what is the current brightness level?''
    \item \textbf{Moderately underspecified commands} allude to desired actions/explanations and relevant settings, but do not name them specifically, e.g., ``make it dimmer'' or ``how bright is that?''
    \item \textbf{Fully underspecified commands} make no reference to device settings or values, but still have an implicit goal, e.g., ``it's too bright'' or ``why is it so bright?''.
\end{itemize}

We assume the ability to infer whether a command is action- or explanation-oriented is a given, i.e., we assume a device can infer whether a command $c$ is either $c_\alpha$ or $c_\epsilon$. Furthermore, we leave user commands with no action- or explanation-oriented goal out of scope, i.e., conversational commands like ``hey lamp, how's it going today?'' have no specified response.

We assume that the \textbf{device state} that is acted upon or explained by a thoughtful thing is finite, and that a formal state model for the device is tractable. This applies to a variety of everyday devices, e.g., lights, thermostats, household appliances, televisions, etc., but does not apply to more complex continuous or real-time ``systems of systems'' like whole automobiles, robots, etc. Furthermore, we do not capture the temporal relationship between subsequent states: the thoughtful lamp cannot respond to explanation-oriented commands like ``what color were you yesterday?'' We describe our approach to state modeling in more detail in Section~\ref{sec:system-modeling}.

% \section{Definitions \& Assumptions}

\section{A Framework for Building Thoughtful Things}
\label{sec:framework}
This section introduces our general-purpose framework for building thoughtful things. We leverage a combination of \emph{formal modeling} of a device, along with \emph{generative modeling} of appropriate actions and explanations (provided by a small generative language model). The former provides a grounded definition of the device's capabilities, while the latter provides the flexibility to respond to diverse user commands. The key advantage of our framework is that the engineering effort is limited to designing a device's system model---we fully automate the process of fine-tuning a complementary generative model to reason about a given device's state in response to action- and explanation-oriented user commands. The framework, depicted in Fig.~\ref{fig:framework}, is summarized as follows:

\begin{figure}
    \centering
    \includegraphics[width=\textwidth]{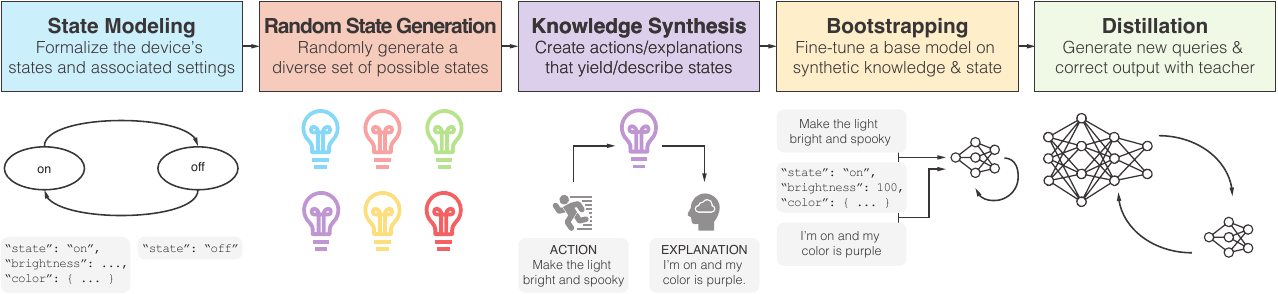}
    \caption{Overview of our framework. Our five-step process leverages a combination of formal modeling, training data synthesis, and fine-tuning and distillation of large language models to train a lightweight model capable of generating appropriate settings and explanations for individual smart devices in response to unconstrained user commands.}
    \label{fig:framework}
\end{figure}

\begin{enumerate}
    \item \textbf{State Modeling.} We first design a formal state model that captures a device's high-level states, as well as the lower-level settings and sensor values that are associated with each state.
    \item \textbf{Random State Generation.} Given a state model of a device, we randomly generate a large set of valid, pseudo-random snapshots of different states to serve as unlabeled training data.
    \item \textbf{Knowledge Synthesis.} For each generated device state snapshot, we use an instruction-tuned teacher LLM to synthesize both an action-oriented command that would result in the device reaching that state, as well as an explanation-oriented command and corresponding explanation that describes that snapshot.
    \item \textbf{Model Bootstrapping.} Using the initial set of synthesized snapshots and annotations, we fine-tune a small (sub-3B-parameter) language model to generate actions and explanations in response to user commands.
    \item \textbf{Model Distillation.} Finally, we refine the initial fine-tuned model by initiating a feedback loop between the bootstrapped ``student'' model and a larger ``teacher'' model that is capable of identifying and remedying inaccuracies in the smaller model's responses.
\end{enumerate}

We describe each of these steps in detail in the following. To aid explanation, we alternate between two motivating examples---a ``thoughtful lamp'' and ``thoughtful thermostat''---that we later implement in Section~\ref{sec:implementations}.

\subsection{State Modeling}
\label{sec:system-modeling}

\begin{figure}
    \centering
    \includegraphics[width=\textwidth]{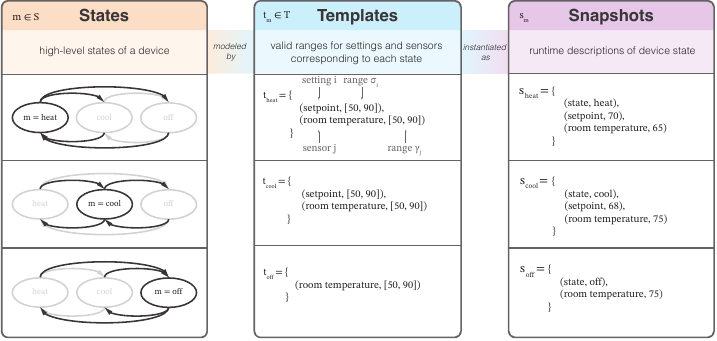}
    \caption{Overview of device state models, with examples included for a thermostat. We model a device based on a high-level state machine $S$ that describes valid transitions between states $m$ (left). A lower-level template $t_m$ associated with each state $m$ captures each setting $i$ and sensor $j$ and their valid ranges $\sigma_i, \gamma_j$ (center). At runtime, a snapshot $s_m$ describes the current state of the device (right). A thoughtful thing leverages a fine-tuned small language model to \emph{act} by generating valid snapshots for new states and to \emph{explain} by describing the snapshot of the device's current state.}
    \label{fig:state}
\end{figure}

\begin{table}[t!]
    \centering
    \begin{tabular}{l|l|p{9cm}} \hline
        & \textbf{Name} & \textbf{Description} \\ \hline \hline
        $\alpha$ & action & machine-parseable description of the settings that a device should have in response to a user command \\ \hline
        $\epsilon$ & explanation & NL description of a device's current state or capabilities \\ \hline
        $c_\alpha$ & action-oriented command & NL command that triggers a device to generate an $\alpha$ \\ \hline
        $c_\epsilon$ & explanation-oriented command & NL command that triggers a device to generate an $\epsilon$ \\ \hline
        % $S$ & state machine & a set of high-level named states $m$ for a device \\
        $m$ & high-level state name & names a high-level device mode or state, e.g., ``on'' state of a light or ``cool'' state of a thermostat \\ \hline
        % $T$ & template set & set of templates $t_m$ for a device\\
        $\sigma_i$ & valid range for setting $i$ & defines the range of possible values for a setting $i$ \\ \hline
        $\Sigma$ & set of valid setting ranges & set of all $\sigma_i$ for a device \\ \hline
        $\gamma_j$ & valid range for sensor input $j$ & defines the range of possible values for a sensor input $j$ \\ \hline
        $\Gamma$ & set of valid sensor input ranges & set of all $\gamma_j$ for a device \\ \hline
        $t_m$ & template for state $m$ & specifies the $(i, \sigma_i)$ and $(j, \gamma_i)$ that are relevant to the state $m$ \\ \hline
        $s_m$ & snapshot of state $m$ & describes the current value $v_i$ of each setting and value $v_j$ of each sensor for a device in state $m$ at runtime \\ \hline
    \end{tabular}
    \caption{Summary of notation we use to formalize the state model of a device. Devices generate either an action $\alpha$ or explanation $\epsilon$ in response to a user command $c$ of the corresponding type. For each state $m$, a corresponding template $t_m$ describes the capabilities of the device in that state. Specifying $t_m$ provides (1)~means to easily generate synthetic data for fine-tuning a language model on a given device's state and (2)~a basis for grounding a language model's actions and explanations in the real capabilities of the device.}
    \label{tab:notation}
\end{table}
The notation introduced in this section (and used throughout the remainder of the paper) is summarized in Table~\ref{tab:notation}. Fig.~\ref{fig:state} provides a helpful visual summary of the state model.

We model a device on two complementary levels of abstraction: (1)~a high-level \emph{state machine} $S$ that expresses the valid transitions between coarse-grained states or ``modes'' $m \in S$ of a device, and (2)~a set $T$ of lower-level \emph{templates} $t_m \in T$ associated with each state $m$. The first step in designing a state model for a device is to identify each of the $m$ that belong to the set of coarse-grained device states $S$. Then, for each $m \in S$ we must design a corresponding $t_m \in T$. A thermostat, for instance, has four high-level states $m \in S = \{{\it fan, heat, cool, off}\}$, and thus four templates $t_m \in T = \{{t_{\it fan}, t_{\it heat}, t_{\it cool}, t_{\it off}\}}$. 

The purpose of a template $t_m$ is to specify the valid ranges for \emph{settings} and, optionally, \emph{sensor inputs} that are \emph{relevant} to the state $m$. We refer to the ``name'' of a setting as $i$ and a sensor input as $j$. We distinguish between settings and sensors in this manner because it allows us to later enforce the fields of a device's state that an underlying generative model is or is not permitted to change. In general, a device has a set $\Sigma$ of individual settings $\sigma_i \in \Sigma$ and a set $\Gamma$ of individual sensor inputs $\gamma_j \in \Gamma$. Both $\sigma_i$ and $\gamma_j$ define valid ranges for a corresponding setting or sensor input with the name $i$ or $j$, e.g., a lamp's percent brightness setting $\sigma_{\it brightness} = [1, 100]$. For, e.g., a thermostat, the device's full set of possible settings is $\Sigma_{\it thermostat} = \{\sigma_{\it setpoint}\}$ and sensor values is $\Gamma_{\it thermostat} = \{\gamma_{\it room\_temperature}\}$. The lamp has settings $\Sigma_{\it lamp} = \{\sigma_{\it brightness}, \sigma_{\it r}, \sigma_{\it g}, \sigma_{\it b}\}$ and no sensors, i.e., $\Gamma_{\it lamp} = \varnothing$.

Each template $t_m$ contains a subset of the $\sigma_i \in \Sigma$ and the $\gamma_j \in \Gamma$, depending on whether the setting or sensor input is relevant to the state $m$. A $\sigma_i$ is relevant to $m$ if $\sigma_i$ influences the behavior of the device in that state. A $\gamma_i$ is relevant to $m$ if (1)~$\gamma_i$ influences the behavior of the device in that state, or (2)~$\gamma_i$ influences the choice of an action $\alpha$ that would result in the device transitioning to a different state $m^\prime$ that is accessible from $m$. We illustrate relevance with a concrete example. The $\sigma_{\it setpoint}$ of a thermostat is relevant to only a subset of states, i.e., when the thermostat is set to either $m = {\it heat}$ or $m = {\it cool}$. In the ${\it fan}$ and ${\it off}$ modes, $\sigma_{\it setpoint}$ is \emph{not} relevant since the setpoint does not influence the operation of the thermostat in those modes. $\sigma_{\it setpoint}$ therefore appears only in $t_{\it heat}$ and $ t_{\it cool}$. The thermostat also has a sensor input $\gamma_{\it room\_temperature}$ in order to regulate the operation of the system. This sensor input is relevant to every state $m \in S$, and thus appears in every template $t_m \in T$. This is because the thermostat needs awareness of the temperature (even when, e.g., in the {\it off} state) in order to choose appropriate future states (i.e., take an action $\alpha$) in response to user commands. For example, if the thermostat is $m = {\it off}$ and the user gives the command $c_\alpha =$ ``it's too hot in here'', the thermostat needs awareness of the room temperature in order to choose an appropriate setpoint as it transitions to the $m^\prime = {\it cool}$ state. In an implementation, $t_m$ takes the form of an associative array (e.g., a Python dictionary), where each relevant setting name $i$ or sensor name $j$ is a key of $t_m$ and the corresponding $\sigma_i$ and $\gamma_i$ is a value:

\begin{equation}
t_m = \{(i, \sigma_i) \mid \sigma_i \in \Sigma, \sigma_i {\rm\ is\ relevant\ to\ } m\} \cup \{(j, \gamma_j) \mid \gamma_j \in \Gamma, \gamma_j {\rm\ is\ relevant\ to\ } m \} \nonumber
\end{equation}
 
At runtime, a device's current state is described by a \emph{snapshot} $s_m$, which is an instantiated template that contains each setting $i$ and sensor input $j$ from the keys of the corresponding template $t_m$, with values $v_i$ and $v_j$ assigned. Additionally, every $s_m$ contains a default ``state'' element that names the high-level state it is associated with, i.e., $\forall s_m, ({\it ``state"}, m) \in s_m$. A snapshot $s_m$ is, like $t_m$, an associative array. A state snapshot is only valid if all $i$ and $j$ in $s_m$ exist in the corresponding $t_m$, and their values are within the ranges specified by $\sigma_i$ and $\gamma_j$:

\begin{equation}
    s_m = \{(i, v_i) \mid \forall i, i \in t_m \wedge v_i \in \sigma_i \} \cup  \{(j, v_j) \mid \forall j, j \in t_m \wedge v_j \in \gamma_j \} \cup \{({\it ``state"}, {\it m}), m \in S\} \nonumber
\end{equation}

State snapshots are important for two reasons. First, they provide a concrete description of the device's current state runtime, from which explanations $\epsilon$ can be generated. When we later fine-tune a small language model to explain the device's state (see Section~\ref{sec:bootstrapping}), we are essentially training it to describe the current snapshot $s_m$ in a manner that satisfies a user's explanation-oriented command, $c_\epsilon$. For example, if a user gives the command $c_\epsilon =$ ``how bright is that?'' and $({\it brightness}, v_{\it brightness} = 50) \in s_m$, the device might respond with $\epsilon = $ ``The brightness is currently set to 50\%''. Furthermore, a state snapshot $s_m$ is nearly interchangeable with a valid action $\alpha$ that our language model should generate in response to an action-oriented user command $c_\alpha$. When we fine-tune it to generate actions, we are training it to generate a form of snapshot that satisfies $c_\alpha$. The only difference between an action $\alpha$ and a full snapshot $s_m$ is that $\alpha$ omits sensor values. The reason is simple: sensor inputs are, by nature, immutable. We define a valid $\alpha$ that transitions the device from a state $m$ to $m^\prime$ as follows:

\begin{equation}
    \alpha = \{(i, v_i) \mid \forall i, i \in t_m \wedge v_i \in \sigma_i \} \cup \{({\it ``state"}, {\it m^\prime}), m^\prime \in S, m^\prime {\rm is\ reachable\ from\ } m\} \nonumber
\end{equation}

Note that the new state $m^\prime$ may equal $m$, since the device need not transition to a different high-level state in order to change its settings. When the device generates a proposed action $\alpha$ in response to a user command $c_\alpha$, we verify that the response adheres to this definition before changing the state of the device. Since we are relying on a generative model to create $\alpha$---and these models are prone to occasional ``hallucination'' by, e.g., creating settings and values that do not exist~\cite{king2024sasha}---the proposed action $\alpha$ occasionally does not satisfy the definition of $\alpha$. When the device generates an invalid $\alpha$, we simply leave the state of the device unchanged.

\subsection{Random State Generation}
We use a given device's state model (specifically, the templates $t_m \in T$) to create a dataset for fine-tuning a small generative language model to act upon and explain the device's state. We begin this process by generating a large number of randomized state snapshots $s_m$ from $t_m$, with the fields of each $s_m$ randomly initialized. The set of random snapshots that we generate needs to be balanced with respect to the relative size of the state spaces of each $m$, as implied by each $t_m \in T$. This is because we want the model to learn information about diverse states (i.e., diverse $s_m$), so that it will generalize well. For example, assume a lamp device has two states, $m \in S = \{{\it on}, {\it off}\}$, where $t_{\it on}$ specifies four relevant settings for brightness and RGB color and $t_{\it off}$ specifies no relevant settings, i.e., $t_{\it off} = \varnothing$. Based on these templates, there is only one possible snapshot in the $m = {\it off}$ state, i.e., $s_{\it off}=\{({\it ``state"}, {\it m})\}$. In the $m = {\it on}$ state, however, there are many combinations of values for the brightness and RGB settings and, correspondingly, a significantly larger number of possible $s_{\it on}$. If we generate random snapshots for $m = {\it on}$ and $m = {\it off}$ with equal probability, our model will be biased toward knowledge about the simple ${\it off}$ state while lacking knowledge about, e.g., the diverse colors expressible in the ${\it on}$ state. We therefore generate snapshots from weighted random choice, taking into account the size of the state space implied by the template for each state. We first make a weighted choice of a state $m \in S$ for which to generate a snapshot with probability $P(m)$. $P(m)$ is simply a function of the number of settings and sensor inputs present in $t_m$:

\begin{equation}
    P(m) = \frac{| \forall i \in t_m | + | \forall j \in t_m | + 1}{|\Sigma \cap \Gamma| + 1}\nonumber
\end{equation}

We add $1$ to the numerator and denominator to account for the fact that a $t_m$ can be empty (e.g., for the lamp in the ${\it off}$ state), but still needs to be chosen with a nonzero probability. After choosing an $m$, we then initialize each value $v_i$ and $v_j$ in $s_m$ uniformly at random from the ranges $\sigma_i$ and $\gamma_j$ specified in $t_m$. This yields a single random snapshot. We repeat this process until we have generated a large set of random snapshots. We determine the appropriate size of this set heuristically, dependent on the size of the state space for a given device. Since the small language model that we ultimately fine-tune on these states incorporates out-of-domain knowledge, it is sufficient to generate a set on the order of hundreds or thousands of state snapshots, even when the state space for, e.g., the thoughtful lamp constitutes billions of possible snapshots.

% \begin{algorithm}[H]
% \SetAlgoLined
% \KwData{List $templates$ of state templates, $N$ sample size}
% \KwResult{List $snapshots$ of randomly-generated state snapshots}
% Initialize weights list $template\_weights$\;
% \For{$i \leftarrow 0$ \KwTo number of state templates in $templates$}{
%     $weight \leftarrow 1$\;
%     Compute a weight for each state based on the number of possible snapshots\;
%     \For{$j \leftarrow 0$ \KwTo number of elements in $templates[i]$}{
%         $weight \leftarrow weight * length(templates[i][j])$
%     }
%     $template\_weights[i] \leftarrow weight$
% }
% \For{$i \leftarrow 0$ \KwTo $N$}{
%     Choose a random state template, weighted by the number of possible snapshots\;
%     $random\_template \leftarrow weighted\_random\_choice(template\_weights, templates)$\;
%     Randomly initialize the values in the template to create a random snapshot\;
%     $random\_snapshot \leftarrow randomly\_initialize\_elements(random\_template)$\;
%     Append $random\_snapshot$ to $snapshots$\;
% }
% \Return $snapshots$\;
% \caption{Random Snapshot Generation}
% \label{alg:random-snapshot}
% \end{algorithm}

\subsection{Knowledge Synthesis}

\begin{figure}
    \centering
    \includegraphics[width=\textwidth]{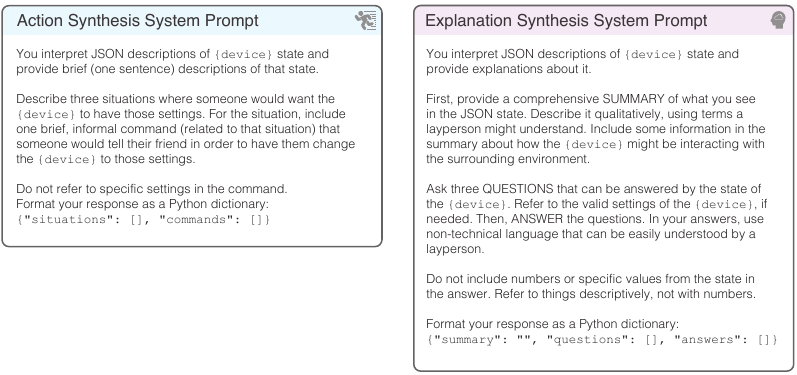}
    \caption{System prompts that we use to synthesize fine-tuning data for small language models. Given a set of state states that we randomly generate using our state model, we use a large instruction-tuned LLM to generate synthetic ``knowledge'' about how different device states relate to different user commands, actions, and explanations.}
    \label{fig:synthesis-prompts}
\end{figure}

We label our set of pseudo-random state snapshots with help from an instruction-tuned LLM, allowing us to create a dataset for fine-tuning a single, smaller generative language model at the action/explanation task for a given device. In our case, the goal of this process is to capture device-specific ``knowledge'' that describes how different user commands $c$, actions $\alpha$, and explanations $\epsilon$ relate to one another. We accomplish this by synthesizing a set of examples of:

\begin{enumerate}
    \item an action-oriented command $c_\alpha$ that yields an action $\alpha$, derived from each random state snapshot $s_m$
    \item an explanation-oriented command $c_\epsilon$ that yields an explanation $\epsilon$ of each random snapshot $s_m$
\end{enumerate}

In the case of (1), the output of the synthesis process is a large set of tuples of $(c_\alpha, \alpha)$, where $\alpha$ contains all of the mutable settings from a random snapshot $s_m$ while omitting any immutable sensor values; the output for (2) is a large set of tuples of $(c_\epsilon, s_m, \epsilon)$. Table~\ref{tab:training-examples} provides examples of random snapshots and synthesized $c_{\alpha}, c_{\epsilon},$ and $\epsilon$. We synthesize instances using two prompts---one for actions, one for explanations---carefully engineered so that they generalize to different devices (Fig.~\ref{fig:synthesis-prompts}).

Our action prompt instructs the teacher LLM to imagine a situation where a user may want the device to have the settings in $\alpha$ given the sensor subset of $s_m$ (if applicable), then create a brief command $c_\alpha$ in informal language that a user would give to the device in that situation. Though instructing the teacher model to imagine a situational context is not strictly necessary to synthesize commands, it tends to produce more loosely-constrained synthetic commands (e.g., ``hey lamp, I'm heading out for the day'') than simply instructing to create a command (which results in, e.g., ``turn off the lamp'')~\cite{prystawski2024think}. We ensure that the set of synthetic $c_\alpha$ are diverse by re-prompting the model to suggest a different command if the semantic similarity (i.e., cosine similarity between word embeddings~\cite{kenter2015short, reimers-2019-sentence-bert}) of the current command and previously-synthesized commands is too high.

To synthesize explanation annotations, we prompt the teacher model to ask and answer a question for each high-level field in the $s_m$, i.e., to synthesize several $c_\epsilon$ and $\epsilon$ for each setting and sensor value in $s_m$. We include the state template $t_m$ in the prompt to ground the teacher's questions and answers in the actual capabilities of the device. For example, if $s_m$ contains $({\it brightness}, 42)$ and the synthesized $c_\epsilon$ is ``Is this the maximum brightness?'', the template field $({\it brightness}, [1, 100])$ suggests a correct $\epsilon$ is, e.g., ``No, this is under half of the maximum brightness.'' When synthesizing instances for a new device, we simply substitute the name of the device (e.g., ``smart light'' or ``thermostat'') into the generic action and explanation prompts with no changes to the instructions.

\begin{table}
    \centering
    \begin{tabular}{l|c|c|c} \hline
         \textbf{Snapshot} $s_m$ & \textbf{Action Command} $c_\alpha$ & \textbf{Explanation Command} $c_\epsilon$ & \textbf{Explanation} $\epsilon$ \\ \hline \hline
         \makecell[l]{\texttt{\{} \\ \texttt{\quad``state'': ``off''} \\ \texttt{\}}} & ``I'm off to bed.'' & \makecell[c]{``Why's it so dark \\ in here?''} &  \makecell[c]{``It's dark because \\ the lamp is off.''} \\ \hline
         \makecell[l]{\texttt{\{} \\ \texttt{\quad``state'': ``on'',} \\ \texttt{\quad``brightness'': 100,} \\ \texttt{\quad``color'':\{} \\ \texttt{\quad\quad``r'': 235,} \\ \texttt{\quad\quad``g'': 64,} \\ \texttt{\quad\quad``b'': 52} \\ \texttt{\quad\}} \\ \texttt{\}}} & \makecell[c]{``Let there be \\ bright red light!''} & \makecell[c]{``Is that as bright \\ as it gets?''} &  \makecell[c]{``Yes, the lamp is \\ at 100\% brightness.''} \\ \hline
    \end{tabular} 
    \caption{Random state snapshots and examples of synthesized commands and explanations for a thoughtful lamp. Given a large set of random snapshots, we use an instruction-tuned teacher LLM to synthesize action- and explanation-oriented commands, as well as explanations. These labeled snapshots serve as a training set for fine-tuning a small LLM to take actions and provide explanations in response to user commands.}
    \label{tab:training-examples}
\end{table}

\subsection{Bootstrapping}
\label{sec:bootstrapping}
\begin{figure}
    \centering
    \includegraphics[width=\textwidth]{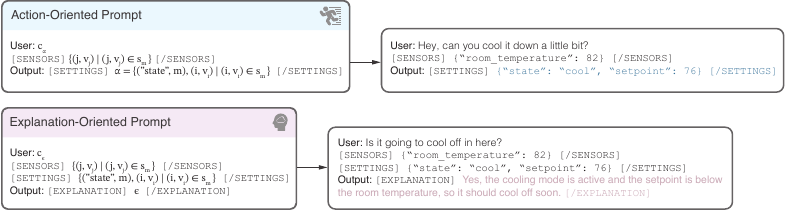}
    \caption{Prompt formats that we use to fine-tune a small language model for device actions/explanations. Examples show prompts for a thermostat device, which has both ``sensors'' and ``settings'' in its state model. When fine-tuning, we substitute the full command, state snapshot (split into immutable sensors and mutable settings), and explanation (if applicable) into the prompt to create training instances. The fine-tuned model learns to complete the prompt with either actions or explanations (depicted as colored text) depending on the prompt.}
    \label{fig:prompts}
\end{figure}

We fine-tune a small language model on our generated snapshots to give it the ability to generate actions $\alpha$ or explanations $\epsilon$ in response to user commands $c_{\alpha}$ or $c_{\epsilon}$. The objective of this ``bootstrapping'' phase is to teach the model the prompt formats for action- and explanation-oriented inputs. We individually insert each of the action tuples $(c_\alpha, \alpha)$ and explanation tuples $(c_\epsilon, s_m, \epsilon)$ from our synthesized dataset into the respective prompt format to create fine-tuning examples (Fig.~\ref{fig:prompts}). By fine-tuning a model on these prompt formats, we are effectively teaching it to complete the input text of the prompt with either an action (i.e., mutable settings) or explanation.

As is conventional, we perform a train-test split of the fine-tuning examples prior to training. However, we hold out the test set beyond the bootstrapping phase and the distillation phase (see Section~\ref{sec:discussion}). This enables us to later test the model's ability to generalize to new commands, beyond simply learning the input-output pairs it has been fine-tuned on. We evaluate models' performance in Section~\ref{sec:evaluation} on this hold-out test set.

\subsection{Distillation}
\label{sec:distillation}
After the first pass of fine-tuning, we use an iterative distillation process to better specialize the small model at the action/explanation task for a device. The goal of this process is to train the language model beyond the point of simply memorizing the prompt structure, into a better ``understanding'' of a given device's capabilities.

\begin{figure}
    \centering
    \includegraphics[width=\textwidth]{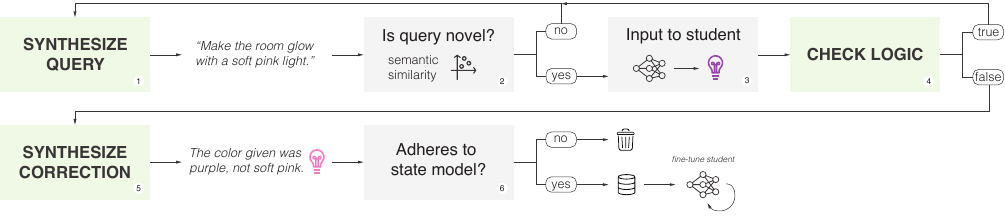}
    \caption{We use an iterative distillation process to align the outputs of the small language model in a thoughtful thing (i.e., the ``student'' model) with the outputs of a higher-performance ``teacher'' LLM. Green boxes denote prompts to the teacher model. This process distills specialized knowledge about a given device's capabilities into the generative model that a thoughtful thing uses to act and explain in response to user commands.}
    \label{fig:distillation}
\end{figure}

Conventional distillation works by minimizing the loss between a ``student'' model's outputs on unlabeled training data and pseudo-labels from a ``teacher'' model on the same data~\cite{buciluǎ2006model}. Recent approaches prioritize synthesis of new training data from the teacher model to further fine-tune the student model, thus ``aligning'' the quality of the student's outputs with those of the teacher~\cite{wang2022self, hsieh2023distilling}. Our approach works in the same spirit as recent approaches, but provides grounding by incorporating the device's state model. We use a teacher LLM to generate progressively more challenging and/or novel action- and explanation-oriented commands to give to the student, and to synthesize the correct outputs to commands that the student is unable to correctly respond to. We illustrate the flow chart for this process in Fig.~\ref{fig:distillation} and describe its steps in the following.

\begin{enumerate}
    \item \textbf{Synthesize query.} We prompt the teacher model to synthesize a new command (either action- or explanation-oriented) to give to the student model. We include the device's state model (i.e., the templates $t_m \in T$) in the prompt to ensure that the command adheres to the actual capabilities of the device.
    \item \textbf{Check novelty.} If the command is action-oriented, we check its novelty against every command the model has been fine-tuned on in the past (including the hold-out test set created during bootstrap fine-tuning). We compute the cosine similarity of the word embeddings of the command against every past command and consider it novel if this similarity is below a threshold. We do not check the novelty of explanation-oriented queries, since the same command may yield different outputs depending on the state of the device.
    \item \textbf{Input to student.} We input the synthesized command to the student model. If the command is action-oriented, we initialize the student to the ``off'' state before inputting; if the command is explanation-oriented, we generate a random snapshot and initialize the device in that state.
    \item \textbf{Check logic.} We prompt the teacher model with the command, the student response, and the full set of templates $T$, instructing it to assign a binary ``true'' or ``false'' label depending on whether the student response satisfies the user's request expressed in the command. If the teacher label is ``true'', we return to~(1). If the rate at which the student responds correctly to the teacher commands exceeds $80\%$, we proceed to~(7) before stopping distillation.
    \item \textbf{Synthesize correction.} If the student responded incorrectly to the teacher's command, we prompt the teacher to synthesize a more appropriate response that the student should have given. We include the device's state model (i.e., templates $t_m \in T$) in the prompt to reduce the likelihood that the teacher synthesizes a response that is not possible given the device's true capabilities.
    \item \textbf{Check adherence to state model.} To prevent fine-tuning the student model on hallucinations produced by the teacher model, we check that the teacher's correction adheres to the state model for the device.
    \item \textbf{Fine-tune.} When the teacher is no longer able to (a)~synthesize commands above the novelty threshold and (b)~consistently produce corrections that adhere to the state model, or (c)~we have generated an equal number of action- and explanation-oriented corrections meeting a preset ``batch size'', we fine-tune the student model on the current batch of synthesized corrections.
\end{enumerate}

After fine-tuning, we repeat the process on the newly distilled version of the student model. We continue until the teacher model is unable to synthesize new commands that are both of sufficient novelty while respecting the state model of the device, i.e., when the rate at which corrections do not adhere to the state model exceeds $90\%$.

\subsection{Integration}
Once we have completed the two main tasks of (1)~designing a device's state model and (2)~fine-tuning a generative model to handle user commands, we integrate these two pieces into the device's system implementation. Given an interface through which to receive commands (e.g., a voice interface) and determine whether a command is action- or explanation-oriented, we prompt the fine-tuned model with the user command and device state snapshot (using the appropriate prompt format in Fig.~\ref{fig:prompts}). If the command elicits an explanation, we extract the explanation text from between the \texttt{[EXPLANATION][/EXPLANATION]} tokens in the model's output; similarly, if the desired output is an action, we extract the new settings for the device from between the \texttt{[SETTINGS][/SETTINGS]}. In the former case, we can display the explanation on a user interface, or read it aloud via text-to-speech. In the latter case, we first attempt to parse the text of the action into a data structure. If parsing fails due to incorrect syntax, we quietly log an error and leave the state of the device unchanged. If successful, we check that the proposed action $\alpha$ adheres to the definition (Sec.~\ref{sec:system-modeling}). If it does not, we quietly log an error without changing the device's settings. If so, we call whichever internal APIs are necessary to change the device's settings to the values implied by the output action $\alpha$.

\section{Case Study Implementations}
\label{sec:implementations}
We use our framework to build two proof-of-concept implementations---a thoughtful lamp and thoughtful thermostat. For each device, we fine-tune two different small language models at the reasoning task---Gemma (2B)~\cite{team2024gemma} and Phi-2 (2.7B)~\cite{gunasekar2023textbooks}---which we later compare in Section~\ref{sec:evaluation}. The engineering effort in all cases is limited to designing a state model for the device, after which the process of synthesizing training data and fine-tuning the generative language model for responding to user commands is automated by our framework. We implement both devices using Python bindings for the llama.cpp framework~\cite{ggerganov2024, abetlen2024}, and therefore adopt Python syntax throughout this section. We run the implementations on a Raspberry Pi 5 (8GB).

\subsection{A Thoughtful Lamp}
The thoughtful lamp is an interesting case because of the size of its state space: in the {\it on} state, the lamp has $1.65 \times 10^9$ possible state snapshots, accounting for each combination of RGB value and brightness. The main task for the lamp's generative model is to reason about the way that these different brightness and color settings relate to different user goals and inquiries. Exploring this case allows us to determine how well a small language model can learn to generalize to the action/explanation task when the size of the dataset we fine-tune it with covers only a very small subset of possible states.

\begin{figure}
    \centering
    \includegraphics[width=0.5\textwidth]{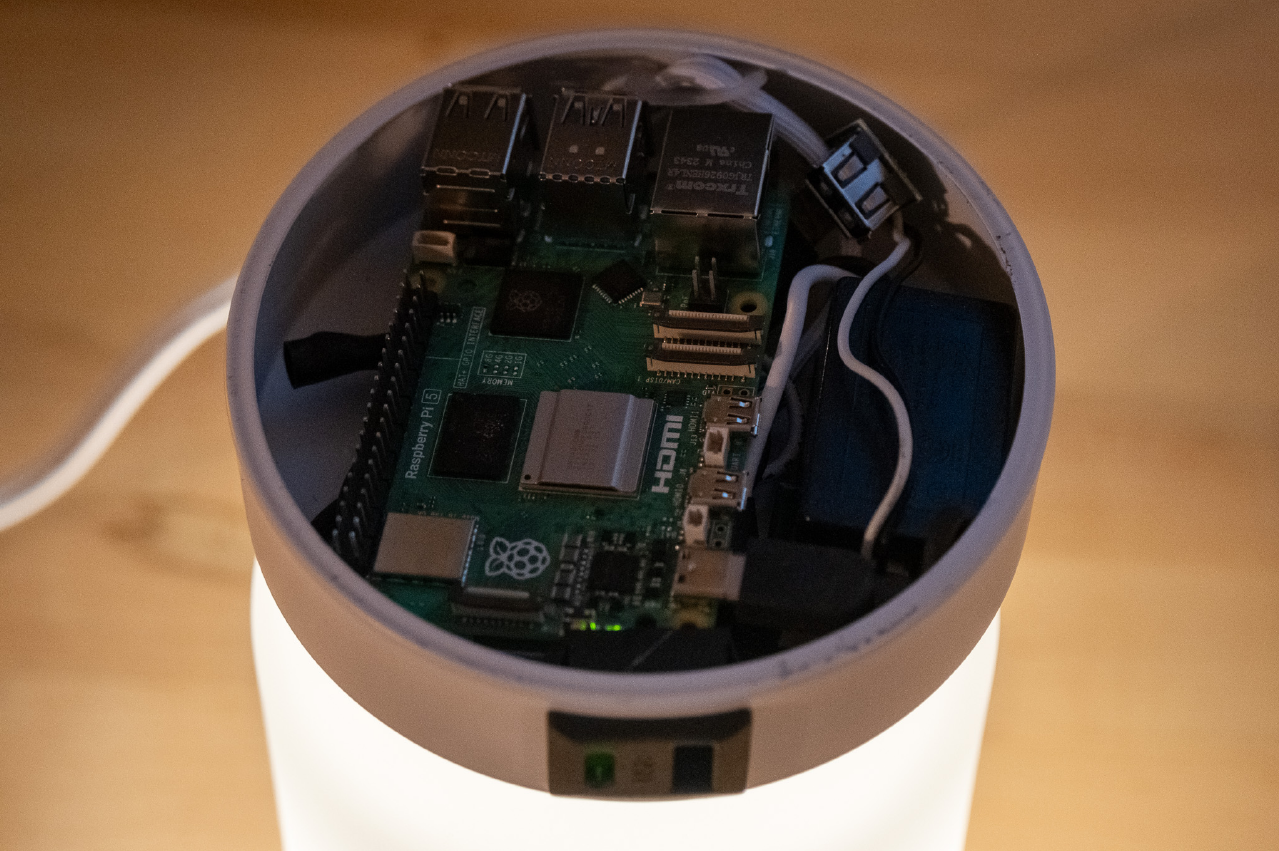}
    \caption{Raspberry Pi 5 installed in the base of our thoughtful lamp prototype. Thoughtful things leverage generative language models for device control and explainability; these models are lightweight enough to run on low-cost commodity hardware.}
    \label{fig:lamp-pi}
\end{figure}

\emph{State Modeling.} Adhering to the definition in Section~\ref{fig:framework}, the thoughtful lamp has two possible high-level states, with $S = \{{\it off, on}\}$. Both states $m \in S$ can transition to one another bidirectionally. The $t_{\it off}$ template contains no settings $\sigma_i$ or sensor inputs $\gamma_j$, i.e., $t_{\it off} = \varnothing$. The template $t_{\it on}$ defines brightness and color settings, expressed as RGB values, i.e., $t_{\it on} = \{\sigma_{\it brightness}, \sigma_{\it r}, \sigma_{\it g}, \sigma_{\it b}\}$. We summarize the state model design in Table~\ref{tab:lamp-state}.

\begin{table}[t!]
    \centering
    \begin{tabular}{l|l|l} \hline
         \textbf{State} $m$ & \textbf{Template} $t_m$ & \textbf{Example Snapshot} $s_m$ \\ \hline \hline
         on &  $t_{\it on} =$ \makecell[l]{\texttt{\{} \\ \texttt{\quad"brightness": range(0, 100),} \\ \texttt{\quad"color": \{} \\ \texttt{\quad\quad"r": range(0, 255),} \\ \texttt{\quad\quad"g": range(0, 255),} \\ \texttt{\quad\quad"b": range(0, 255)} \\ \texttt{\quad\}} \\ \texttt{\}}} & $s_{\it on} =$\makecell[l]{\texttt{\{} \\ \texttt{\quad"state": "on",} \\ \texttt{\quad"brightness": 55,} \\ \texttt{\quad"color": \{} \\ \texttt{\quad\quad"r": 64,} \\ \texttt{\quad\quad"g": 234,} \\ \texttt{\quad\quad"b": 1} \\ \texttt{\quad\}} \\ \texttt{\}}} \\ \hline
        % Action & \multicolumn{2}{l}{
        %  \makecell[l]{
        %     User: Hey, can you adjust the lights to a soft purple to match the party theme? \\
        %     Output: \text{[SETTINGS]} \texttt{\{"state": "on", "brightness": 22, "color": \{"r": 175, "g": 124, "b": 186\}\}} \text{[/SETTINGS]}
        %  }} \\ \hline
        % Explanation & \multicolumn{2}{l}{
        %  \makecell[l]{
        %     User: What is the brightness level of the smart light? \\ 
        %     \text{[SETTINGS]} \texttt{\{"state": "on", "brightness": 22, "color": \{"r": 175, "g": 124, "b": 186\}\}} \text{[/SETTINGS]} \\
        %     Output: [EXPLANATION] The smart light's brightness is not at its maximum, but it\'s also not very dim. It is somewhere in between, closer to the dimmer side. [/EXPLANATION]
        %  }} \\ \hline
         off & $t_{\it off} =$ \texttt{\{\}} & $s_{\it off} =$ \makecell[l]{\texttt{\{} \\ \texttt{\quad"state": "off"} \\ \texttt{\}}} \\ \hline
    \end{tabular}
    \caption{Summary of state model for the thoughtful lamp (expressed as Python dictionaries). Left: two high-level states $m \in S = \{{\it on, off}\}$; Center: templates $t_m$ for each state; Right: examples of snapshots $s_m$ for each state. The lamp has four setting values in the on state (brightness and RGB color) and no sensor inputs; in the off state, the lamp has no settings.}
    \label{tab:lamp-state}
\end{table}

\emph{Random State Generation \& Knowledge Synthesis.} We use our state model to randomly generate 200 states, then synthesize their corresponding annotations using GPT-4. We generate two annotations per state---one action-oriented and one explanation-oriented---resulting in 400 total fine-tuning instances. 

\emph{Bootstrapping \& Distillation.} We perform bootstrap fine-tuning on a small language model with our initial set of synthetic instances. We do this using low-rank adaptation (LoRA)~\cite{hu2021lora}, which keeps the requirement for GPU VRAM below the 16GB provided by the machine instance we perform training on. After bootstrapping is complete, we merge the LoRA adapter layer into the full model weights and begin distillation. We perform distillation with GPT-4 as the teacher model. This process takes several hours, during which the student model receives around $1000$ commands that are synthesized by GPT-4. This produces several hundred new training instances based on GPT-4's corrections, which we then fine-tune the student model on. Both models required a comparable number of corrections during distillation.

\begin{figure}
    \centering
    \includegraphics[width=0.8\textwidth]{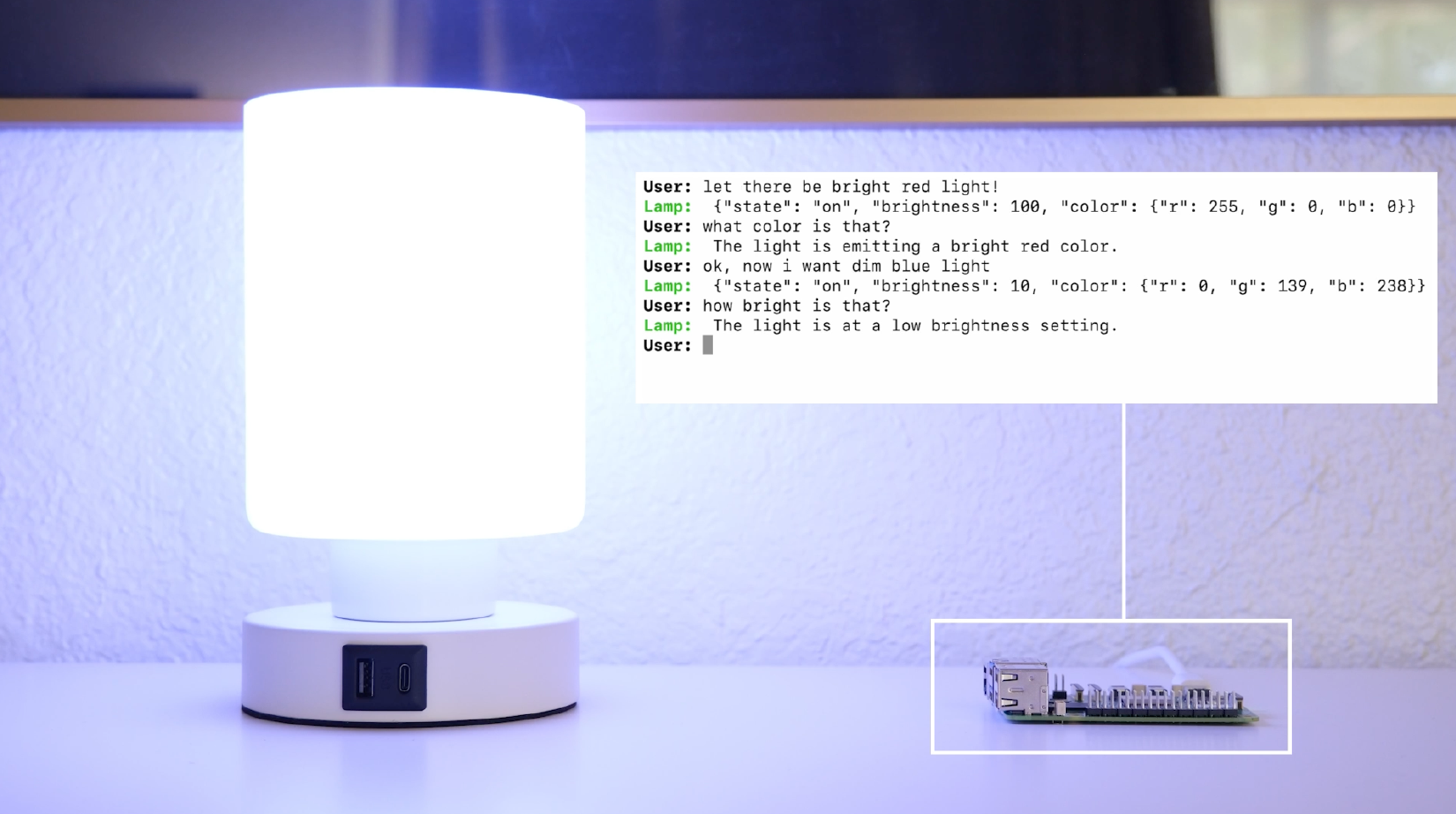}
    \caption{Example interaction with the thoughtful lamp, via command line interface (Pi removed from base for illustration).}
    \label{fig:lamp-convo}
\end{figure}

\emph{Integration.} We embed the Raspberry Pi in the base of the lamp (Fig.~\ref{fig:lamp-pi}), which controls a Philips Hue smart bulb. While a speech-to-text interface is possible using, e.g., a lightweight Whisper model~\cite{radford2023robust}, we rely on a command line interface for simplicity. When the language model generates new actions, we parse them as JSON before mapping the changes to settings to the appropriate Philips Hue API calls. Explanations output by the model are simply displayed on the command line. We show an example of several interactions in Fig.~\ref{fig:lamp-convo}.

\subsection{A Thoughtful Thermostat}
The thoughtful thermostat has a smaller state space than the lamp, but incorporates sensor input (i.e., room temperature) in its state model. We undertake this implementation to determine how well a small language model can learn to differentiate between mutable and immutable state (i.e., setting and sensor values, respectively).

\emph{State Modeling.} The thermostat has four possible states, with $S = \{{\it fan, heat, cool, off}\}$. Each state $m \in S$ can transition to each other state.  In only the ${\it heat}$ and ${\it cool}$ modes, the thermostat has a $\sigma_{\it setpoint}$ setting; the thermostat has a $\gamma_{\it room\_temperature}$ sensor input in every state. We summarize the state model design in Table~\ref{tab:thermostat-state}.

\begin{table}[t!]
    \centering
    \begin{tabular}{l|l|l} \hline
         \textbf{State} $m$ & \textbf{Template} $t_m$ & \textbf{Example Snapshot} $s_m$ \\ \hline \hline
         heat &  $t_{\it heat} =$ \makecell[l]{\texttt{\{} \\ \texttt{\quad"room\_temperature": range(50, 90),} \\ \texttt{\quad"setpoint": range(50, 90)} \\ \texttt{\}}} 
         & $s_{\it heat} =$ \makecell[l]{\texttt{\{} \\ \texttt{\quad"state": "heat",} \\ \texttt{\quad"room\_temperature": 55,} \\ \texttt{\quad"setpoint": 68} \\ \texttt{\}}} \\ \hline
         cool & $t_{\it cool} =$ \makecell[l]{\texttt{\{} \\ \texttt{\quad"room\_temperature": range(50, 90),} \\ \texttt{\quad"setpoint": range(50, 90)} \\ \texttt{\}}} 
         & $s_{\it cool} =$ \makecell[l]{\texttt{\{} \\ \texttt{\quad"state": "cool",} \\ \texttt{\quad"room\_temperature": 81,} \\ \texttt{\quad"setpoint": 75} \\ \texttt{\}}} \\ \hline
         fan & $t_{\it fan} =$ \makecell[l]{\texttt{\{} \\ \texttt{\quad"room\_temperature": range(50, 90)} \\ \texttt{\}}} & $s_{\it fan} =$ \makecell[l]{\texttt{\{} \\ \texttt{\quad"state": "fan",} \\ \texttt{\quad"room\_temperature": 74} \\ \texttt{\}}} \\ \hline
         off & $t_{\it off} =$ \makecell[l]{\texttt{\{} \\ \texttt{\quad"room\_temperature": range(50, 90)} \\ \texttt{\}}} & $s_{\it off} =$ \makecell[l]{\texttt{\{} \\ \texttt{\quad"state": "off",} \\ \texttt{\quad"room\_temperature": 74} \\ \texttt{\}}} \\
         \hline
    \end{tabular}
    \caption{Summary of state model for the thoughtful thermostat (expressed as Python dictionaries). Left: four high-level states $m \in S = \{{\it heat, cool, fan, off}\}$; Center: templates $t_m$ for each state; Right: examples of snapshots $s_m$ for each state. The thermostat has one sensor input (room temperature), which is relevant to every state and appears in every template. In heat and cool modes, the thermostat has a setpoint setting.}
    \label{tab:thermostat-state}
\end{table}

\emph{Random State Generation.} We randomly generate 200 states, then synthesize their corresponding action/explanation annotations using GPT-4, resulting in 400 bootstrap fine-tuning instances.

\emph{Bootstrapping \& Distillation.} As before, we perform bootstrap fine-tuning with our synthetic instances, aided by LoRA on a 16GB VRAM machine. After bootstrapping, we use GPT-4 as a teacher model for distillation. Unlike the lamp, the two base models behaved differently from one another during distillation for the thermostat: Phi-2 required one long round before meeting the threshold to end distillation, while Gemma required two short rounds. Many of the Gemma model's responses during the first distillation round were erroneous, suggesting that the model was still learning the prompt format during distillation. The model therefore quickly met the threshold for corrections to end the first round. Both models received around $500$ novel commands from the teacher model before completing distillation, which is about half that of the lamp. % This may suggest that the thermostat's state space is easier to learn than the lamp's.

We also noted that both models began to learn actions that assigned only a subset of settings available for a state (e.g., changing the setpoint of the thermostat without changing the mode). While these actions are valid, this behavior was not anticipated---all of our bootstrap fine-tuning instances contain the full set of settings for each state. We attribute the differences between the distillation of the lamp and thermostat to differences in prompt structure (the thermostat prompt includes sensors, while the lamp's does not).

\emph{Integration.} We install the Raspberry Pi on the rear of a 7" touchscreen display, which provides a user interface to the implementation. We input commands via keyboard, though a speech-to-text implementation is also feasible. When the language model generates actions or explanations in response to a command, we display them on the screen. We do not connect our implementation to a climate control system---however, the Pi provides general-purpose input/output (GPIO) pins capable of interfacing with a common 4-wire residential system, making this extra integration step fairly trivial.

\section{Evaluation}
\label{sec:evaluation}
We frame our evaluations of our implementations around the following research questions:

\begin{itemize}
    \item \emph{\textbf{RQ1:} How well do small language models perform at the device action/explanation task when trained in our framework?} Performing ``well'' in this case means that a model gains the ability to generate actions and explanations that set and explain a given device's state and settings with high accuracy.
    \item \emph{\textbf{RQ2:} Can small language models learn to generalize to the action/explanation task through fine-tuning?} Training a model on every possible state it could encounter is not feasible. Ideally, we can leverage the out-of-domain knowledge present in a language model to generalize to new and unseen states given only a small set of training examples.
    \item \emph{\textbf{RQ3:} What is the practical performance when implemented in a real device?} Since our long-term objective is to integrate ``thoughtfulness'' into everyday devices, it is important to understand runtime performance, e.g., response latency.
\end{itemize}

We investigate these research questions using the procedures described in the following.

\subsection{Procedures \& Metrics}
For each implementation, we use our framework to fine-tune two base models at the action/explanation task: Gemma (2B)~\cite{team2024gemma} and Phi-2 (2.7B)~\cite{gunasekar2023textbooks}. We use a hold-out test set of $100$ commands (evenly split between action-oriented and explanation-oriented), created during the bootstrapping phase of fine-tuning (see Section~\ref{sec:bootstrapping}). We input each command from our hold-out test set, generating either an action or explanation with the model under test. For action-oriented commands, we initialize the device's state model to $m = {\it off}$. For explanation-oriented commands, we initialize the device with a randomly-generated snapshot, with the random seed held constant between trials so that the set of states explained by each respective model are the same. Toward answering \textbf{RQ1}, we measure performance along the following metric: 

\begin{itemize}
    \item \textbf{Setting-level accuracy.} For each generated response, we use GPT-4 to assign a ``correct'' or ``incorrect'' label, indicating whether or not the response was appropriate given the command (and device state, if applicable). For responses with an ``incorrect'' label, we use GPT-4 to label which of the fields in the device's state snapshot were either (1)~incorrectly set (for actions) or (2)~incorrectly explained (for explanations). We perform this for the initial bootstrapped model and the distilled model, each at two levels of precision---16-bit floating point (i.e., fp16) and 8-bit integer (int8) quantization. To strengthen our confidence in GPT-4's labels, we validate them with a human annotation task performed by 3 annotators. For each response generated by the fp16 distilled models, annotators provide a binary ``correct'' or ``incorrect'' label. Each annotation instance shows the command given and the two models' responses. We instruct the annotator to select each output that correctly addresses the command, or ``none of the above'' if none were correct. This produces $600$ labels with Krippendorf's $\alpha = 0.75$, suggesting satisfactory agreement between human annotators~\cite{hayes2007answering}. We then check agreement between each human annotator and GPT-4, yielding pairwise Cohen's $\kappa$ of $0.66$, $0.76$, and $0.73$, respectively, suggesting substantial agreement~\cite{landis1977measurement}.
\end{itemize}

Toward answering \textbf{RQ2}, we compare the similarity of the models' fine-tuning inputs to the models' generated outputs on the hold-out test set using the following metrics:

\begin{itemize}
    \item \textbf{Jaccard similarity.} Action-oriented commands generate outputs with high syntactic overlap, since they must adhere to the device's state model. We thus measure the portion of output actions that are fully identical to those in the training set using Jaccard set similarity. Lower Jaccard similarity suggests that a model is generating novel snapshots with setting values not seen in the training set. We exclude {\it off} states from this metric, since there is only one possible form a snapshot can take for this state.
    \item \textbf{ROUGE score.} For explanation-oriented commands, we measure the ROUGE score~\cite{lin2004rouge} between the synthesized explanations and the explanations output by the model. Lower scores indicate lower similarity. In our case, lower is better since it suggests that the model is generating novel explanations.
\end{itemize}

Toward answering \textbf{RQ3}, we deploy each distilled model at two levels of quantization (fp16 and int8) on our Raspberry Pi 5 (8GB). We do not utilize any form of GPU acceleration. We use our set of $100$ test commands as input to collect the following metrics:

\begin{itemize}
    \item \textbf{Tokens per second.} We measure the overall speed of responses by dividing the number of tokens in the output by the time taken to respond.
    \item \textbf{Response latency by interaction type.} We measure the time taken to respond, as a function of the command type (action- or explanation-oriented).
    \item \textbf{Memory usage.} We measure the fixed amount of (non-GPU) memory usage by each model at runtime.
    \item \textbf{CPU temperature.} Intensive operations like LLM inferences produce heat, which results in CPU throttling and reduced performance. We evaluate this impact by continuously inputting our test commands over an hour span and measuring the change in temperature. We also note whether any CPU throttling occurs during this span. During these trials, we place the Raspberry Pi 5 in a case with thermal heat sink padding.
\end{itemize}

\subsection{Results}
\begin{figure}
    \centering
    \includegraphics[width=\textwidth]{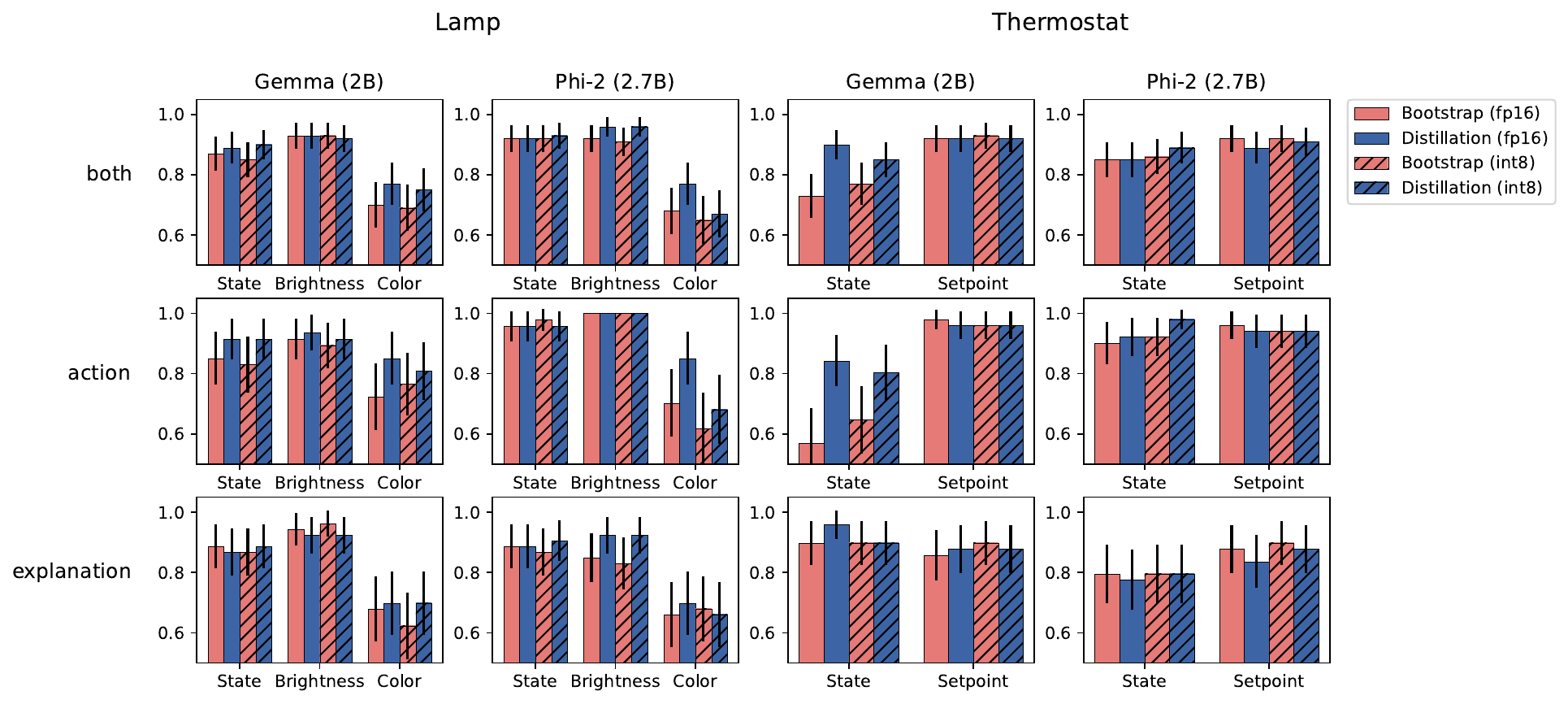}
    \caption{Model accuracy at choosing and/or explaining different device settings across responses to 100 test commands. Top: both command types; middle: action-oriented commands; bottom: explanation-oriented commands. If a bar shows, e.g., 80\% for ``brightness'', this means that 80\% of responses appropriately chose and/or explained the device's brightness setting in response to a user command. In general, small state spaces are learned more quickly, performance in larger state spaces benefits from more training, and explanations are harder to learn than actions. Error bars depict a 90\% confidence interval.}
    \label{fig:scores_all}
\end{figure}

\emph{{\textbf{Models quickly learn correct actions and explanations for small state spaces.}}} Fig.~\ref{fig:scores_all} depicts the accuracy at choosing or explaining the given device's settings, across model responses. For example, if the accuracy for ``brightness'' is 80\%, this means that 80\% of the model's responses chose or explained the device's brightness correctly. In general, models quickly attain high performance at choosing and/or explaining settings that have small state spaces, regardless of device or command type. For the lamp, the number of choices for ``state'' is two (${\it on}$ or ${\it off}$); all models achieve over 80\% accuracy at either acting appropriately when the correct response is to turn the light on or off, or correctly explaining the state (e.g., when $c_\epsilon = $ ``Is the lamp using electricity?'', generating $\epsilon =$ ``No, the lamp is not using electricity because it is off"). This is also true of the lamp's brightness, and the setpoint of the thermostat. As mentioned in Section~\ref{sec:implementations}, the thermostat models met the threshold to complete distillation after about half as many student/teacher iterations as the lamp, meaning the models are trained on an overall smaller set of instances---these models nonetheless achieve performance comparable to the lamp.

\emph{{\textbf{Distillation improves the quality of actions and explanations for larger state spaces.}}} While models can quickly achieve high performance for small state spaces, larger state spaces pose more of a challenge. Referring again to Fig.~\ref{fig:scores_all}, the performance for ``color'' is lowest among all models, for all command types. However, distillation in these cases appears to provide the largest improvement in performance over the bootstrapped models. This suggests that additional training is most beneficial to models' performance at the task of choosing/explaining states like ``color'' that represent very large decision spaces. We note that Gemma appears at first glance to struggle with the relatively small state space of the thermostat ``state'' setting (i.e., {\it on, off, heat, cool, fan}), and improves with distillation. This is not because the state space itself is difficult---rather, the bootstrapped model produced a large number of invalid outputs since it had not fully learned the prompt structure. 

\emph{\textbf{Models generate novel \& valid actions and explanations not seen in the training set.}} Table~\ref{tab:similarity} depicts the textual similarity between models' outputs and their training data at successive phases of training in our framework. For actions, the Jaccard similarity is zero for all models at the bootstrap phase, meaning none of the models' outputs are identical to those in the synthetic training set of actions. This number increases only slightly after distillation---the models begin to memorize a very small number of states with more fine-tuning. For explanations, the ROUGE scores decrease with more training---the score for each ROUGE metric is at its lowest for every model after distillation. This suggests that more training via distillation induces the models to generate a more diverse set of explanations. In conjunction with the results in Fig.~\ref{fig:scores_all}, the models' generated outputs are not simply novel, but also exhibit a reasonable level of accuracy. With respect to \textbf{RQ2}, models do not aggressively memorize the synthetic actions and explanations they are fine-tuned on---rather, they learn to generate novel actions and explanations.

\begin{table}[t!]
\begin{tabular}{l|l|l|lllll}
\hline
\textbf{Device} & \textbf{Base Model} & \textbf{Checkpoint} & \textbf{Jaccard}     & \textbf{ROUGE-1}     & \textbf{ROUGE-2}     & \textbf{ROUGE-L}     \\
\hline \hline
\multirow{4}{*}{Lamp}       & \multirow{2}{*}{Gemma}    & bootstrap    & \textbf{0.00} & 0.28          & 0.13          & 0.26          \\
                            &               & distillation & 0.01          & \textbf{0.21} & \textbf{0.06} & \textbf{0.19} \\ \cline{2-7}
                            & \multirow{2}{*}{Phi-2}  & bootstrap    & \textbf{0.00} & 0.35          & 0.16          & 0.32          \\
                            &               & distillation & 0.05          & \textbf{0.30} & \textbf{0.12} & \textbf{0.27} \\ \hline
\multirow{4}{*}{Thermostat} & \multirow{2}{*}{Gemma}    & bootstrap    & \textbf{0.00} & 0.39          & 0.14          & 0.34          \\
                            &               & distillation & 0.04          & \textbf{0.27} & \textbf{0.09} & \textbf{0.23} \\ \cline{2-7}
                            & \multirow{2}{*}{Phi-2}  & bootstrap    & \textbf{0.00} & 0.41          & 0.14          & 0.35          \\
                            &               & distillation & 0.04          & \textbf{0.33} & \textbf{0.10} & \textbf{0.28} \\
\hline
\end{tabular}
    \caption{Similarity between models' generated actions/explanations and the actions/explanations in the training set. All models are fp16 precision. Jaccard measures similarity between training and output actions; ROUGE scores measure similarity between training and output explanations. In general, the models do not aggressively memorize the training set---rather, they learn to generate novel actions and explanations in response to user commands.}
    \label{tab:similarity}
\end{table}

\emph{\textbf{Latency, memory usage, and temperature in the base case hint at feasibility in practice.}} Table~\ref{tab:empirical-pi} reports results for each device, base model, and precision when evaluated on our Raspberry Pi 5. Our implementation does not fully leverage the hardware capabilities of the Pi (e.g., GPU acceleration), nor do we use memory locking to ensure that the models are fully allocated in RAM. In general, the quantized models provide greater responsiveness. The thermostat device generally exhibits lower latency than the lamp since its state structure requires fewer tokens. When giving each model a continuous stream of commands for an hour, we noted that the temperature of the Raspberry Pi never reached the $80^{\circ}C$ threshold for CPU throttling. Temperature tended to increase by about $5^{\circ}C$ to $10^{\circ}C$ over the first 15 minutes before stabilizing. With respect to \textbf{RQ3}, the base implementation with no platform-specific optimizations has moderate responsiveness and does not suffer from performance degradation under load.

% \begin{table}[t!]
% \begin{tabular}{l|l|l|llll}
% \hline
% \textbf{Device} & \textbf{Base Model} & \textbf{Precision} & \textbf{Tokens/s} & $\alpha$ \textbf{Latency (s)} & $\epsilon$ \textbf{Latency (s)} & \textbf{RAM} \\ 
% \hline \hline 
% \multirow{4}{*}{Lamp} & \multirow{2}{*}{Gemma (2B)}   & fp16 & 0.96 & 31.36 & 41.01 & \\
%                       &                               & int8 & & & & \\ \cline{2-7}
%                       & \multirow{2}{*}{Phi-2 (2.7B)} & fp16 & 0.97 & 25.72 & 37.31 & \\
%                       &                               & int8 & 2.37 & 10.44 & 14.81 & \\ \hline
% \multirow{4}{*}{Thermostat} & \multirow{2}{*}{Gemma (2B)}   & fp16 & 0.53 & 25.47 & 38.57 & \\
%                             &                               & int8 & 1.20 & 10.15 & 14.81 & \\ \cline{2-7}
%                             & \multirow{2}{*}{Phi-2 (2.7B)} & fp16 & 0.54 & 26.84 & 43.16 & \\
%                             &                               & int8 & 1.31 & 10.59 & 17.67 & \\ \hline
% \end{tabular}
%     \caption{Empirical results of runtime testing on a Raspberry Pi 5 (using llama.cpp framework without GPU acceleration).}
%     \label{tab:empirical-pi}
% \end{table}

\begin{table}[t!]
\centering
    \begin{tabular}{l|l|l|c|c|rrr|rrr}
     \multicolumn{5}{l}{} &   \multicolumn{3}{c}{\textbf{$\alpha$ Latency (s)}} &   \multicolumn{3}{c}{\textbf{$\epsilon$ Latency (s)}}  \\ \hline
    \textbf{Device}       & \textbf{Model}   & \textbf{Prec.}   &  \textbf{Mem. (GB)} & \textbf{Tokens/s} & Mean & Min & Max & Mean & Min & Max \\
    
    \hline \hline

     \multirow{4}{*}{Lamp}              & \multirow{2}{*}{Gemma}        & fp16   &  5.81    &       0.60 & 36.51 & 10.06 & 73.56    & 27.52 & 18.58  & 48.54       \\
                                        &                               & int8   &  3.15    &       1.25 & 16.94 & 5.20  & 34.50    & 13.50 & 8.98   & 23.15       \\ \cline{2-11}
                                        & \multirow{2}{*}{Phi-2}        & fp16   &  5.51    &       0.63 & 25.11 & 9.34  & 33.96    & 22.49 & 12.22  & 33.04       \\
                                        &                               & int8   &  3.08    &       1.13 & 13.35 & 5.29  & 17.96    & 13.07 & 5.97   & 17.73       \\ \hline
     \multirow{4}{*}{Thermostat}        & \multirow{2}{*}{Gemma}        & fp16   &     &       0.80 & 18.95 & 13.29 & 23.51    & 21.95 & 13.70  & 30.39       \\
                                        &                               & int8   &     &       1.51 & 9.68  & 6.02  & 13.24    & 11.30 & 7.90   & 14.68       \\ \cline{2-11}
                                        & \multirow{2}{*}{Phi-2}        & fp16   &     &       0.68 & 20.47 & 15.42 & 24.46    & 26.56 & 21.47  & 33.20       \\
                                        &                               & int8   &     &       1.24 & 11.65 & 8.63  & 13.94    & 14.43 & 10.47  & 18.01       \\
    \hline
    \end{tabular}
    \caption{Empirical results of runtime testing on a Raspberry Pi 5 using the Python bindings for the llama.cpp framework (with no GPU acceleration or memlocking). Without memory locking (i.e., storing the full model in memory without paging any parts to disk), memory usage results reflect a combination of RAM and disk usage. 8-bit quantized models are generally faster than 16-bit models, with lower memory overhead. Average latency for actions is generally higher for the lamp than for the thermostat, since the light's state model requires more tokens. In general, high latency motivates more specialized implementations that make optimal use of system resources (rather than general-purpose inference frameworks)}
    \label{tab:empirical-pi}
\end{table}

\section{Discussion}
\label{sec:discussion}
Our framework instills small language models with ``self-knowledge'' about a device, which enables the models to generate appropriate actions and explanations in response to user commands. These language models are small enough to integrate on real devices, with no need for cloud infrastructure. We achieve this without a need for specialized training data---by simply designing a state model for a device, we are able to automate the rest of the process. Our effort opens the door for several further avenues of research, which we discuss in the following.
 
\emph{\textbf{Mastering massive state spaces.}} Our results suggest that larger state spaces are harder for a model to learn. While our general-purpose framework handles the process of adapting a generalist model to action/explanation for a device, models would likely benefit from additional domain-specific knowledge in the training set. An important takeaway is that \emph{some device states are harder to learn than others}. Bearing this in mind when building a specialized training set, model performance would likely benefit from data that elucidates those broader states. For the lamp, this would be, e.g., a large set of named RGB colors and unconstrained text describing them~\cite{eiseman2017complete}. Training new base models on high-quality, domain-specific knowledge would likely provide added benefit to downstream performance~\cite{gunasekar2023textbooks}.

\emph{\textbf{Improving runtime responsiveness.}} Our empirical results suggest that practical thoughtful things are within reach on everyday hardware, but we leave room for improvement with respect to response latency. The first step toward improving an implementation is to integrate it more closely with the capabilities of the platform---e.g., to utilize GPU acceleration and avoid memory paging. We noted during development that responses using the llama.cpp binaries tended to be about 3x faster than using the open-source Python bindings for llama.cpp, suggesting that closer integration with the framework would provide performance gains. In a production implementation, the hardware platform and software implementation of the state model and of generative model inferences are likely to be tightly coupled, providing higher performance. 

\emph{\textbf{Incorporating temporality of state.}} Our state model uses snapshots to encode immediate states, and to describe future states. Given the limitations of small language models, we stop short of incorporating a temporal dimension to state, where a sequence of device behaviors over time can be explained or proposed as an action. This extension to our framework would nonetheless benefit user-centricity. Users are often confused by, e.g., smart devices that execute learned (or user-programmed but forgotten) automation routines unpredictably~\cite{he2019smart}. While our model allows for a device to explain its current state relative to possible future states (e.g., the lamp at 50\% brightness can be made brighter), it has no knowledge of past decisions or states. The ability to schedule future states, or to transition through sequences of states would also be beneficial. Interestingly, many of the ``teacher'' corrections that we observed for the lamp during distillation involved such plans: GPT-4 would create a command like ``cycle through the colors of the rainbow'', then synthesize JSON suggesting this state sequence. These corrections, however, do not adhere to our current state model, so they were omitted.

\section{Conclusion}
\label{sec:conclusion}
In this paper, we proposed \emph{thoughtful things}: human-centric devices empowered by a combination of formal state models and generative language models to take action and provide explanations in response to unconstrained user commands. We described our framework for building thoughtful things, which takes a state model for a given device and uses it to synthesize data and fine-tune a small generative language model for the action/explanation task. The models that enable thoughtful things' reasoning are lightweight enough to run on-device, with no dependency on the cloud. To demonstrate our framework, we undertook two case study implementations---a thoughtful lamp and thoughtful thermostat. We evaluated our implementations by performing automated and human annotation of the quality of models' responses. Finally, we deployed our implementations on real hardware (a Raspberry Pi 5) and collected empirical measurements of response latency, memory consumption, and computational demand over time. Our work opens the door for everyday devices with greater usability and improved user privacy.

% \begin{acks}
% \end{acks}

\bibliographystyle{ACM-Reference-Format}
\bibliography{refs}


\begin{thebibliography}{64}

%%% the Web address.

\ifx \showCODEN    \undefined \def \showCODEN     #1{\unskip}     \fi
\ifx \showDOI      \undefined \def \showDOI       #1{#1}\fi
\ifx \showISBNx    \undefined \def \showISBNx     #1{\unskip}     \fi
\ifx \showISBNxiii \undefined \def \showISBNxiii  #1{\unskip}     \fi
\ifx \showISSN     \undefined \def \showISSN      #1{\unskip}     \fi
\ifx \showLCCN     \undefined \def \showLCCN      #1{\unskip}     \fi
\ifx \shownote     \undefined \def \shownote      #1{#1}          \fi
\ifx \showarticletitle \undefined \def \showarticletitle #1{#1}   \fi
\ifx \showURL      \undefined \def \showURL       {\relax}        \fi
% invisible to TeX
\providecommand\bibfield[2]{#2}
\providecommand\bibinfo[2]{#2}
\providecommand\natexlab[1]{#1}
\providecommand\showeprint[2][]{arXiv:#2}

\bibitem[Achiam et~al\mbox{.}(2023)]%
        {achiam2023gpt}
\bibfield{author}{\bibinfo{person}{Josh Achiam}, \bibinfo{person}{Steven Adler}, \bibinfo{person}{Sandhini Agarwal}, \bibinfo{person}{Lama Ahmad}, \bibinfo{person}{Ilge Akkaya}, \bibinfo{person}{Florencia~Leoni Aleman}, \bibinfo{person}{Diogo Almeida}, \bibinfo{person}{Janko Altenschmidt}, \bibinfo{person}{Sam Altman}, \bibinfo{person}{Shyamal Anadkat}, {et~al\mbox{.}}} \bibinfo{year}{2023}\natexlab{}.
\newblock \showarticletitle{GPT-4 Technical Report}.
\newblock \bibinfo{journal}{\emph{arXiv preprint arXiv:2303.08774}} (\bibinfo{year}{2023}).
\newblock


\bibitem[Appliances(2023a)]%
        {ge2023nonsmart}
\bibfield{author}{\bibinfo{person}{GE Appliances}.} \bibinfo{year}{2023}\natexlab{a}.
\newblock \bibinfo{booktitle}{\emph{Owner's Manual, Refrigerator Models 16, 17, 18, 19, 22} (\bibinfo{edition}{6th} ed.)}.
\newblock General Electric.
\newblock


\bibitem[Appliances(2023b)]%
        {ge2023smart}
\bibfield{author}{\bibinfo{person}{GE Appliances}.} \bibinfo{year}{2023}\natexlab{b}.
\newblock \bibinfo{booktitle}{\emph{Owner's Manual, Refrigerator Models PVD, PXD, and GVE} (\bibinfo{edition}{7th} ed.)}.
\newblock General Electric.
\newblock


\bibitem[Ashworth(2022)]%
        {ashworth_smart_nodate}
\bibfield{author}{\bibinfo{person}{Boone Ashworth}.} \bibinfo{year}{2022}\natexlab{}.
\newblock \showarticletitle{The {Risk} of {Relying} on a {Smart} {Home} {Company} to {Keep} {The} {Lights} {On}}.
\newblock \bibinfo{journal}{\emph{Wired}} (\bibinfo{date}{April} \bibinfo{year}{2022}).
\newblock
\showISSN{1059-1028}
\urldef\tempurl%
\url{https://www.wired.com/story/insteon-shutdown/}
\showURL{%
\tempurl}


\bibitem[Barbosa et~al\mbox{.}(2020)]%
        {barbosa2020privacy}
\bibfield{author}{\bibinfo{person}{Nat{\~a}~M Barbosa}, \bibinfo{person}{Zhuohao Zhang}, {and} \bibinfo{person}{Yang Wang}.} \bibinfo{year}{2020}\natexlab{}.
\newblock \showarticletitle{Do Privacy and Security Matter to Everyone? Quantifying and Clustering $\{$User-Centric$\}$ Considerations About Smart Home Device Adoption}. In \bibinfo{booktitle}{\emph{Sixteenth Symposium on Usable Privacy and Security (SOUPS 2020)}}. \bibinfo{pages}{417--435}.
\newblock


\bibitem[Bell(2023)]%
        {bell_erg_nodate}
\bibfield{author}{\bibinfo{person}{Tom Bell}.} \bibinfo{year}{2023}\natexlab{}.
\newblock \bibinfo{title}{{ERG} mode explained: what it is, how to use it and when you should turn it off}.
\newblock
\newblock
\urldef\tempurl%
\url{https://www.bikeradar.com/advice/fitness-and-training/erg-mode}
\showURL{%
\tempurl}


\bibitem[Betlen(2024)]%
        {abetlen2024}
\bibfield{author}{\bibinfo{person}{Andrei Betlen}.} \bibinfo{year}{2024}\natexlab{}.
\newblock \bibinfo{title}{llama-cpp-python}.
\newblock \bibinfo{howpublished}{\url{https://github.com/abetlen/llama-cpp-python}}.
\newblock


\bibitem[Brade et~al\mbox{.}(2024)]%
        {brade2024synthscribe}
\bibfield{author}{\bibinfo{person}{Stephen Brade}, \bibinfo{person}{Bryan Wang}, \bibinfo{person}{Mauricio Sousa}, \bibinfo{person}{Gregory~Lee Newsome}, \bibinfo{person}{Sageev Oore}, {and} \bibinfo{person}{Tovi Grossman}.} \bibinfo{year}{2024}\natexlab{}.
\newblock \showarticletitle{SynthScribe: Deep Multimodal Tools for Synthesizer Sound Retrieval and Exploration}. In \bibinfo{booktitle}{\emph{Proceedings of the 29th International Conference on Intelligent User Interfaces}}. \bibinfo{pages}{51--65}.
\newblock


\bibitem[Buciluǎ et~al\mbox{.}(2006)]%
        {buciluǎ2006model}
\bibfield{author}{\bibinfo{person}{Cristian Buciluǎ}, \bibinfo{person}{Rich Caruana}, {and} \bibinfo{person}{Alexandru Niculescu-Mizil}.} \bibinfo{year}{2006}\natexlab{}.
\newblock \showarticletitle{Model compression}. In \bibinfo{booktitle}{\emph{Proceedings of the 12th ACM SIGKDD international conference on Knowledge discovery and data mining}}. \bibinfo{pages}{535--541}.
\newblock


\bibitem[Burmeister et~al\mbox{.}(2015)]%
        {burmeister2015ambient}
\bibfield{author}{\bibinfo{person}{Daniel Burmeister}, \bibinfo{person}{Bashar Altakrouri}, {and} \bibinfo{person}{Andreas Schrader}.} \bibinfo{year}{2015}\natexlab{}.
\newblock \showarticletitle{Ambient Reflection: Towards Self-explaining Devices.}. In \bibinfo{booktitle}{\emph{LMIS@ EICS}}. \bibinfo{pages}{16--20}.
\newblock


\bibitem[Burmeister et~al\mbox{.}(2017)]%
        {burmeister2017smart}
\bibfield{author}{\bibinfo{person}{Daniel Burmeister}, \bibinfo{person}{Florian Burmann}, {and} \bibinfo{person}{Andreas Schrader}.} \bibinfo{year}{2017}\natexlab{}.
\newblock \showarticletitle{The smart object description language: modeling interaction capabilities for self-reflection}. In \bibinfo{booktitle}{\emph{2017 IEEE International Conference on Pervasive Computing and Communications Workshops (PerCom Workshops)}}. IEEE, \bibinfo{pages}{503--508}.
\newblock


\bibitem[Camp(2019)]%
        {camp_my_nodate}
\bibfield{author}{\bibinfo{person}{Jeffrey~Van Camp}.} \bibinfo{year}{2019}\natexlab{}.
\newblock \showarticletitle{My {Jibo} {Is} {Dying} and {It}'s {Breaking} {My} {Heart}}.
\newblock \bibinfo{journal}{\emph{Wired}} (\bibinfo{date}{March} \bibinfo{year}{2019}).
\newblock
\showISSN{1059-1028}
\urldef\tempurl%
\url{https://www.wired.com/story/jibo-is-dying-eulogy/}
\showURL{%
\tempurl}


\bibitem[Ciuffoletti(2018)]%
        {ciuffoletti2018low}
\bibfield{author}{\bibinfo{person}{Augusto Ciuffoletti}.} \bibinfo{year}{2018}\natexlab{}.
\newblock \showarticletitle{Low-cost IoT: A holistic approach}.
\newblock \bibinfo{journal}{\emph{Journal of Sensor and Actuator Networks}} \bibinfo{volume}{7}, \bibinfo{number}{2} (\bibinfo{year}{2018}), \bibinfo{pages}{19}.
\newblock


\bibitem[Coppers et~al\mbox{.}(2020)]%
        {coppers2020fortniot}
\bibfield{author}{\bibinfo{person}{Sven Coppers}, \bibinfo{person}{Davy Vanacken}, {and} \bibinfo{person}{Kris Luyten}.} \bibinfo{year}{2020}\natexlab{}.
\newblock \showarticletitle{FORTNIoT: Intelligible Predictions to Improve User Understanding of Smart Home Behavior}.
\newblock \bibinfo{journal}{\emph{Proc. ACM Interact. Mob. Wearable Ubiquitous Technol.}} \bibinfo{volume}{4}, \bibinfo{number}{4}, Article \bibinfo{articleno}{124} (\bibinfo{date}{dec} \bibinfo{year}{2020}), \bibinfo{numpages}{24}~pages.
\newblock
\urldef\tempurl%
\url{https://doi.org/10.1145/3432225}
\showDOI{\tempurl}


\bibitem[Coskun et~al\mbox{.}(2018)]%
        {coskun2018smart}
\bibfield{author}{\bibinfo{person}{Aykut Coskun}, \bibinfo{person}{G{\"u}l Kaner}, {and} \bibinfo{person}{{\.I}dil Bostan}.} \bibinfo{year}{2018}\natexlab{}.
\newblock \showarticletitle{Is smart home a necessity or a fantasy for the mainstream user? A study on users’ expectations of smart household appliances}.
\newblock \bibinfo{journal}{\emph{International Journal of Design}} \bibinfo{volume}{12}, \bibinfo{number}{1} (\bibinfo{year}{2018}), \bibinfo{pages}{7--20}.
\newblock


\bibitem[Cowan et~al\mbox{.}(2017)]%
        {cowan2017can}
\bibfield{author}{\bibinfo{person}{Benjamin~R Cowan}, \bibinfo{person}{Nadia Pantidi}, \bibinfo{person}{David Coyle}, \bibinfo{person}{Kellie Morrissey}, \bibinfo{person}{Peter Clarke}, \bibinfo{person}{Sara Al-Shehri}, \bibinfo{person}{David Earley}, {and} \bibinfo{person}{Natasha Bandeira}.} \bibinfo{year}{2017}\natexlab{}.
\newblock \showarticletitle{" What can i help you with?" infrequent users' experiences of intelligent personal assistants}. In \bibinfo{booktitle}{\emph{Proceedings of the 19th international conference on human-computer interaction with mobile devices and services}}. \bibinfo{pages}{1--12}.
\newblock


\bibitem[Dey(2009)]%
        {dey2009explanations}
\bibfield{author}{\bibinfo{person}{Anind~K Dey}.} \bibinfo{year}{2009}\natexlab{}.
\newblock \showarticletitle{Explanations in Context-Aware Systems.}. In \bibinfo{booktitle}{\emph{ExaCt}}. \bibinfo{pages}{84--93}.
\newblock


\bibitem[Eiseman(2017)]%
        {eiseman2017complete}
\bibfield{author}{\bibinfo{person}{Leatrice Eiseman}.} \bibinfo{year}{2017}\natexlab{}.
\newblock \bibinfo{booktitle}{\emph{The complete color harmony, pantone edition: expert color information for professional results}}.
\newblock \bibinfo{publisher}{Rockport Publishers}.
\newblock


\bibitem[Ferrero et~al\mbox{.}(2019)]%
        {ferrero2019ubiquitous}
\bibfield{author}{\bibinfo{person}{Renato Ferrero}, \bibinfo{person}{Mohammad~Ghazi Vakili}, \bibinfo{person}{Edoardo Giusto}, \bibinfo{person}{Mauro Guerrera}, {and} \bibinfo{person}{Vincenzo Randazzo}.} \bibinfo{year}{2019}\natexlab{}.
\newblock \showarticletitle{Ubiquitous fridge with natural language interaction}. In \bibinfo{booktitle}{\emph{2019 IEEE International Conference on RFID Technology and Applications (RFID-TA)}}. IEEE, \bibinfo{pages}{404--409}.
\newblock


\bibitem[Gerganov(2024)]%
        {ggerganov2024}
\bibfield{author}{\bibinfo{person}{Georgi Gerganov}.} \bibinfo{year}{2024}\natexlab{}.
\newblock \bibinfo{title}{llama.cpp}.
\newblock \bibinfo{howpublished}{\url{https://github.com/ggerganov/llama.cpp}}.
\newblock


\bibitem[Google(2024)]%
        {googlenest}
\bibfield{author}{\bibinfo{person}{Google}.} \bibinfo{year}{2024}\natexlab{}.
\newblock \bibinfo{title}{Beginner's guide to the {N}est thermostat}.
\newblock
\newblock
\urldef\tempurl%
\url{https://support.google.com/googlenest/answer/9248184}
\showURL{%
\tempurl}


\bibitem[Gudovskiy et~al\mbox{.}(2019)]%
        {gudovskiy2019smart}
\bibfield{author}{\bibinfo{person}{Denis Gudovskiy}, \bibinfo{person}{Gyuri Han}, \bibinfo{person}{Takuya Yamaguchi}, {and} \bibinfo{person}{Sotaro Tsukizawa}.} \bibinfo{year}{2019}\natexlab{}.
\newblock \showarticletitle{Smart home appliances: Chat with your fridge}.
\newblock \bibinfo{journal}{\emph{arXiv preprint arXiv:1912.09589}} (\bibinfo{year}{2019}).
\newblock


\bibitem[Gunasekar et~al\mbox{.}(2023)]%
        {gunasekar2023textbooks}
\bibfield{author}{\bibinfo{person}{Suriya Gunasekar}, \bibinfo{person}{Yi Zhang}, \bibinfo{person}{Jyoti Aneja}, \bibinfo{person}{Caio C{\'e}sar~Teodoro Mendes}, \bibinfo{person}{Allie Del~Giorno}, \bibinfo{person}{Sivakanth Gopi}, \bibinfo{person}{Mojan Javaheripi}, \bibinfo{person}{Piero Kauffmann}, \bibinfo{person}{Gustavo de Rosa}, \bibinfo{person}{Olli Saarikivi}, {et~al\mbox{.}}} \bibinfo{year}{2023}\natexlab{}.
\newblock \showarticletitle{Textbooks are all you need}.
\newblock \bibinfo{journal}{\emph{arXiv preprint arXiv:2306.11644}} (\bibinfo{year}{2023}).
\newblock


\bibitem[Hayes and Krippendorff(2007)]%
        {hayes2007answering}
\bibfield{author}{\bibinfo{person}{Andrew~F Hayes} {and} \bibinfo{person}{Klaus Krippendorff}.} \bibinfo{year}{2007}\natexlab{}.
\newblock \showarticletitle{Answering the call for a standard reliability measure for coding data}.
\newblock \bibinfo{journal}{\emph{Communication methods and measures}} \bibinfo{volume}{1}, \bibinfo{number}{1} (\bibinfo{year}{2007}), \bibinfo{pages}{77--89}.
\newblock


\bibitem[He et~al\mbox{.}(2019)]%
        {he2019smart}
\bibfield{author}{\bibinfo{person}{Weijia He}, \bibinfo{person}{Jesse Martinez}, \bibinfo{person}{Roshni Padhi}, \bibinfo{person}{Lefan Zhang}, {and} \bibinfo{person}{Blase Ur}.} \bibinfo{year}{2019}\natexlab{}.
\newblock \showarticletitle{When smart devices are stupid: negative experiences using home smart devices}. In \bibinfo{booktitle}{\emph{2019 IEEE Security and Privacy Workshops (SPW)}}. IEEE, \bibinfo{pages}{150--155}.
\newblock


\bibitem[Hsieh et~al\mbox{.}(2023)]%
        {hsieh2023distilling}
\bibfield{author}{\bibinfo{person}{Cheng-Yu Hsieh}, \bibinfo{person}{Chun-Liang Li}, \bibinfo{person}{Chih-Kuan Yeh}, \bibinfo{person}{Hootan Nakhost}, \bibinfo{person}{Yasuhisa Fujii}, \bibinfo{person}{Alexander Ratner}, \bibinfo{person}{Ranjay Krishna}, \bibinfo{person}{Chen-Yu Lee}, {and} \bibinfo{person}{Tomas Pfister}.} \bibinfo{year}{2023}\natexlab{}.
\newblock \showarticletitle{Distilling step-by-step! outperforming larger language models with less training data and smaller model sizes}.
\newblock \bibinfo{journal}{\emph{arXiv preprint arXiv:2305.02301}} (\bibinfo{year}{2023}).
\newblock


\bibitem[Hu et~al\mbox{.}(2021)]%
        {hu2021lora}
\bibfield{author}{\bibinfo{person}{Edward~J Hu}, \bibinfo{person}{Yelong Shen}, \bibinfo{person}{Phillip Wallis}, \bibinfo{person}{Zeyuan Allen-Zhu}, \bibinfo{person}{Yuanzhi Li}, \bibinfo{person}{Shean Wang}, \bibinfo{person}{Lu Wang}, {and} \bibinfo{person}{Weizhu Chen}.} \bibinfo{year}{2021}\natexlab{}.
\newblock \showarticletitle{Lora: Low-rank adaptation of large language models}.
\newblock \bibinfo{journal}{\emph{arXiv preprint arXiv:2106.09685}} (\bibinfo{year}{2021}).
\newblock


\bibitem[Iancu and Iancu(2020)]%
        {iancu2020love}
\bibfield{author}{\bibinfo{person}{Ioana Iancu} {and} \bibinfo{person}{Bogdan Iancu}.} \bibinfo{year}{2020}\natexlab{}.
\newblock \showarticletitle{I love it, but it is too complicated. Aging adults’ perspective on mobile technology acceptance}.
\newblock \bibinfo{journal}{\emph{ESSACHESS--Journal for Communication Studies}} \bibinfo{volume}{13}, \bibinfo{number}{2 (26)} (\bibinfo{year}{2020}), \bibinfo{pages}{13--39}.
\newblock


\bibitem[Kenter and De~Rijke(2015)]%
        {kenter2015short}
\bibfield{author}{\bibinfo{person}{Tom Kenter} {and} \bibinfo{person}{Maarten De~Rijke}.} \bibinfo{year}{2015}\natexlab{}.
\newblock \showarticletitle{Short text similarity with word embeddings}. In \bibinfo{booktitle}{\emph{Proceedings of the 24th ACM international on conference on information and knowledge management}}. \bibinfo{pages}{1411--1420}.
\newblock


\bibitem[King et~al\mbox{.}(2023)]%
        {king2023get}
\bibfield{author}{\bibinfo{person}{Evan King}, \bibinfo{person}{Haoxiang Yu}, \bibinfo{person}{Sangsu Lee}, {and} \bibinfo{person}{Christine Julien}.} \bibinfo{year}{2023}\natexlab{}.
\newblock \showarticletitle{"Get ready for a party": Exploring smarter smart spaces with help from large language models}.
\newblock \bibinfo{journal}{\emph{arXiv preprint arXiv:2303.14143}} (\bibinfo{year}{2023}).
\newblock


\bibitem[King et~al\mbox{.}(2024)]%
        {king2024sasha}
\bibfield{author}{\bibinfo{person}{Evan King}, \bibinfo{person}{Haoxiang Yu}, \bibinfo{person}{Sangsu Lee}, {and} \bibinfo{person}{Christine Julien}.} \bibinfo{year}{2024}\natexlab{}.
\newblock \showarticletitle{Sasha: creative goal-oriented reasoning in smart homes with large language models}.
\newblock \bibinfo{journal}{\emph{Proceedings of the ACM on Interactive, Mobile, Wearable and Ubiquitous Technologies}} \bibinfo{volume}{8}, \bibinfo{number}{1} (\bibinfo{year}{2024}), \bibinfo{pages}{1--38}.
\newblock


\bibitem[Kojima et~al\mbox{.}(2022)]%
        {kojima2022large}
\bibfield{author}{\bibinfo{person}{Takeshi Kojima}, \bibinfo{person}{Shixiang~Shane Gu}, \bibinfo{person}{Machel Reid}, \bibinfo{person}{Yutaka Matsuo}, {and} \bibinfo{person}{Yusuke Iwasawa}.} \bibinfo{year}{2022}\natexlab{}.
\newblock \showarticletitle{Large language models are zero-shot reasoners}.
\newblock \bibinfo{journal}{\emph{Advances in neural information processing systems}}  \bibinfo{volume}{35} (\bibinfo{year}{2022}), \bibinfo{pages}{22199--22213}.
\newblock


\bibitem[Kordts et~al\mbox{.}(2021)]%
        {kordts2021towards}
\bibfield{author}{\bibinfo{person}{B{\"o}rge Kordts}, \bibinfo{person}{Bennet Gerlach}, {and} \bibinfo{person}{Andreas Schrader}.} \bibinfo{year}{2021}\natexlab{}.
\newblock \showarticletitle{Towards self-explaining ambient applications}. In \bibinfo{booktitle}{\emph{Proceedings of the 14th PErvasive Technologies Related to Assistive Environments Conference}}. \bibinfo{pages}{383--390}.
\newblock


\bibitem[Kulesza et~al\mbox{.}(2013)]%
        {kulesza2013too}
\bibfield{author}{\bibinfo{person}{Todd Kulesza}, \bibinfo{person}{Simone Stumpf}, \bibinfo{person}{Margaret Burnett}, \bibinfo{person}{Sherry Yang}, \bibinfo{person}{Irwin Kwan}, {and} \bibinfo{person}{Weng-Keen Wong}.} \bibinfo{year}{2013}\natexlab{}.
\newblock \showarticletitle{Too much, too little, or just right? Ways explanations impact end users' mental models}. In \bibinfo{booktitle}{\emph{2013 IEEE Symposium on visual languages and human centric computing}}. IEEE, \bibinfo{pages}{3--10}.
\newblock


\bibitem[Kwon and Lim(2017)]%
        {kwon2017multi}
\bibfield{author}{\bibinfo{person}{Sook-Youn Kwon} {and} \bibinfo{person}{Jae-Hyun Lim}.} \bibinfo{year}{2017}\natexlab{}.
\newblock \showarticletitle{Multi-objective context-adaptive natural lighting system}.
\newblock \bibinfo{journal}{\emph{Energy and Buildings}}  \bibinfo{volume}{144} (\bibinfo{year}{2017}), \bibinfo{pages}{61--73}.
\newblock


\bibitem[Landis and Koch(1977)]%
        {landis1977measurement}
\bibfield{author}{\bibinfo{person}{J~Richard Landis} {and} \bibinfo{person}{Gary~G Koch}.} \bibinfo{year}{1977}\natexlab{}.
\newblock \showarticletitle{The measurement of observer agreement for categorical data}.
\newblock \bibinfo{journal}{\emph{biometrics}} (\bibinfo{year}{1977}), \bibinfo{pages}{159--174}.
\newblock


\bibitem[Lazar et~al\mbox{.}(2015)]%
        {lazar2015we}
\bibfield{author}{\bibinfo{person}{Amanda Lazar}, \bibinfo{person}{Christian Koehler}, \bibinfo{person}{Theresa~Jean Tanenbaum}, {and} \bibinfo{person}{David~H Nguyen}.} \bibinfo{year}{2015}\natexlab{}.
\newblock \showarticletitle{Why we use and abandon smart devices}. In \bibinfo{booktitle}{\emph{Proceedings of the 2015 ACM international joint conference on pervasive and ubiquitous computing}}. \bibinfo{pages}{635--646}.
\newblock


\bibitem[Li et~al\mbox{.}(2009)]%
        {li2009intelligent}
\bibfield{author}{\bibinfo{person}{Bojun Li}, \bibinfo{person}{Piyanuch Hathaipontaluk}, {and} \bibinfo{person}{Suhuai Luo}.} \bibinfo{year}{2009}\natexlab{}.
\newblock \showarticletitle{Intelligent oven in smart home environment}. In \bibinfo{booktitle}{\emph{2009 international conference on research challenges in computer science}}. IEEE, \bibinfo{pages}{247--250}.
\newblock


\bibitem[Li et~al\mbox{.}(2021)]%
        {li2021motivations}
\bibfield{author}{\bibinfo{person}{Wenda Li}, \bibinfo{person}{Tan Yigitcanlar}, \bibinfo{person}{Isil Erol}, {and} \bibinfo{person}{Aaron Liu}.} \bibinfo{year}{2021}\natexlab{}.
\newblock \showarticletitle{Motivations, barriers and risks of smart home adoption: From systematic literature review to conceptual framework}.
\newblock \bibinfo{journal}{\emph{Energy Research \& Social Science}}  \bibinfo{volume}{80} (\bibinfo{year}{2021}), \bibinfo{pages}{102211}.
\newblock


\bibitem[Lieberman and Espinosa(2006)]%
        {lieberman2006goal}
\bibfield{author}{\bibinfo{person}{Henry Lieberman} {and} \bibinfo{person}{Jos{\'e} Espinosa}.} \bibinfo{year}{2006}\natexlab{}.
\newblock \showarticletitle{A goal-oriented interface to consumer electronics using planning and commonsense reasoning}. In \bibinfo{booktitle}{\emph{Proceedings of the 11th international conference on Intelligent User Interfaces}}. \bibinfo{pages}{226--233}.
\newblock


\bibitem[Lin(2004)]%
        {lin2004rouge}
\bibfield{author}{\bibinfo{person}{Chin-Yew Lin}.} \bibinfo{year}{2004}\natexlab{}.
\newblock \showarticletitle{ROUGE: A package for automatic evaluation of summaries}. In \bibinfo{booktitle}{\emph{Text summarization branches out}}. \bibinfo{pages}{74--81}.
\newblock


\bibitem[Luger and Sellen(2016)]%
        {luger2016like}
\bibfield{author}{\bibinfo{person}{Ewa Luger} {and} \bibinfo{person}{Abigail Sellen}.} \bibinfo{year}{2016}\natexlab{}.
\newblock \showarticletitle{" Like Having a Really Bad PA" The Gulf between User Expectation and Experience of Conversational Agents}. In \bibinfo{booktitle}{\emph{Proceedings of the 2016 CHI conference on human factors in computing systems}}. \bibinfo{pages}{5286--5297}.
\newblock


\bibitem[Meinicke et~al\mbox{.}(2020)]%
        {meinicke2020multi}
\bibfield{author}{\bibinfo{person}{Lucian Meinicke}, \bibinfo{person}{Keiko Ochi}, {and} \bibinfo{person}{Yasunari Obuchi}.} \bibinfo{year}{2020}\natexlab{}.
\newblock \showarticletitle{Multi-level Query Analysis for NLT-based Synthesizer Interface}. In \bibinfo{booktitle}{\emph{2020 Nicograph International (NicoInt)}}. IEEE, \bibinfo{pages}{62--65}.
\newblock


\bibitem[Parab(2024)]%
        {parab_google_2024}
\bibfield{author}{\bibinfo{person}{Pranay Parab}.} \bibinfo{year}{2024}\natexlab{}.
\newblock \bibinfo{title}{Google {Assistant} {Is} {Losing} a {Bunch} of {Features}}.
\newblock
\newblock
\urldef\tempurl%
\url{https://lifehacker.com/tech/google-assistant-is-losing-a-bunch-of-features}
\showURL{%
\tempurl}
\newblock
\shownote{Section: Internet}.


\bibitem[Paul et~al\mbox{.}(2024)]%
        {paul2024enabling}
\bibfield{author}{\bibinfo{person}{Sudipta Paul}, \bibinfo{person}{Lingyu Zhang}, \bibinfo{person}{Yilin Shen}, {and} \bibinfo{person}{Hongxia Jin}.} \bibinfo{year}{2024}\natexlab{}.
\newblock \showarticletitle{Enabling Device Control Planning Capabilities of Small Language Model}. In \bibinfo{booktitle}{\emph{ICASSP 2024-2024 IEEE International Conference on Acoustics, Speech and Signal Processing (ICASSP)}}. IEEE, \bibinfo{pages}{12066--12070}.
\newblock


\bibitem[Pradhan et~al\mbox{.}(2020)]%
        {pradhan2020use}
\bibfield{author}{\bibinfo{person}{Alisha Pradhan}, \bibinfo{person}{Amanda Lazar}, {and} \bibinfo{person}{Leah Findlater}.} \bibinfo{year}{2020}\natexlab{}.
\newblock \showarticletitle{Use of intelligent voice assistants by older adults with low technology use}.
\newblock \bibinfo{journal}{\emph{ACM Transactions on Computer-Human Interaction (TOCHI)}} \bibinfo{volume}{27}, \bibinfo{number}{4} (\bibinfo{year}{2020}), \bibinfo{pages}{1--27}.
\newblock


\bibitem[Pradhan et~al\mbox{.}(2018)]%
        {pradhan2018accessibility}
\bibfield{author}{\bibinfo{person}{Alisha Pradhan}, \bibinfo{person}{Kanika Mehta}, {and} \bibinfo{person}{Leah Findlater}.} \bibinfo{year}{2018}\natexlab{}.
\newblock \showarticletitle{"Accessibility Came by Accident" Use of Voice-Controlled Intelligent Personal Assistants by People with Disabilities}. In \bibinfo{booktitle}{\emph{Proceedings of the 2018 CHI Conference on human factors in computing systems}}. \bibinfo{pages}{1--13}.
\newblock


\bibitem[Prystawski et~al\mbox{.}(2024)]%
        {prystawski2024think}
\bibfield{author}{\bibinfo{person}{Ben Prystawski}, \bibinfo{person}{Michael Li}, {and} \bibinfo{person}{Noah Goodman}.} \bibinfo{year}{2024}\natexlab{}.
\newblock \showarticletitle{Why think step by step? Reasoning emerges from the locality of experience}.
\newblock \bibinfo{journal}{\emph{Advances in Neural Information Processing Systems}}  \bibinfo{volume}{36} (\bibinfo{year}{2024}).
\newblock


\bibitem[Pyae and Joelsson(2018)]%
        {aung2018}
\bibfield{author}{\bibinfo{person}{Aung Pyae} {and} \bibinfo{person}{Tapani~N. Joelsson}.} \bibinfo{year}{2018}\natexlab{}.
\newblock \showarticletitle{Investigating the usability and user experiences of voice user interface: a case of Google home smart speaker}. In \bibinfo{booktitle}{\emph{Proceedings of the 20th International Conference on Human-Computer Interaction with Mobile Devices and Services Adjunct}} (Barcelona, Spain) \emph{(\bibinfo{series}{MobileHCI '18})}. \bibinfo{publisher}{Association for Computing Machinery}, \bibinfo{address}{New York, NY, USA}, \bibinfo{pages}{127–131}.
\newblock
\showISBNx{9781450359412}
\urldef\tempurl%
\url{https://doi.org/10.1145/3236112.3236130}
\showDOI{\tempurl}


\bibitem[Radford et~al\mbox{.}(2023)]%
        {radford2023robust}
\bibfield{author}{\bibinfo{person}{Alec Radford}, \bibinfo{person}{Jong~Wook Kim}, \bibinfo{person}{Tao Xu}, \bibinfo{person}{Greg Brockman}, \bibinfo{person}{Christine McLeavey}, {and} \bibinfo{person}{Ilya Sutskever}.} \bibinfo{year}{2023}\natexlab{}.
\newblock \showarticletitle{Robust speech recognition via large-scale weak supervision}. In \bibinfo{booktitle}{\emph{International Conference on Machine Learning}}. PMLR, \bibinfo{pages}{28492--28518}.
\newblock


\bibitem[Reimers and Gurevych(2019)]%
        {reimers-2019-sentence-bert}
\bibfield{author}{\bibinfo{person}{Nils Reimers} {and} \bibinfo{person}{Iryna Gurevych}.} \bibinfo{year}{2019}\natexlab{}.
\newblock \showarticletitle{Sentence-BERT: Sentence Embeddings using Siamese BERT-Networks}. In \bibinfo{booktitle}{\emph{Proceedings of the 2019 Conference on Empirical Methods in Natural Language Processing}}. \bibinfo{publisher}{Association for Computational Linguistics}.
\newblock
\urldef\tempurl%
\url{http://arxiv.org/abs/1908.10084}
\showURL{%
\tempurl}


\bibitem[Sadeghi et~al\mbox{.}(2024)]%
        {sadeghi2024smartex}
\bibfield{author}{\bibinfo{person}{Mersedeh Sadeghi}, \bibinfo{person}{Lars Herbold}, \bibinfo{person}{Max Unterbusch}, {and} \bibinfo{person}{Andreas Vogelsang}.} \bibinfo{year}{2024}\natexlab{}.
\newblock \showarticletitle{SmartEx: A Framework for Generating User-Centric Explanations in Smart Environments}.
\newblock \bibinfo{journal}{\emph{arXiv preprint arXiv:2402.13024}} (\bibinfo{year}{2024}).
\newblock


\bibitem[Shier(2021)]%
        {shier2021synthesizer}
\bibfield{author}{\bibinfo{person}{Jordie Shier}.} \bibinfo{year}{2021}\natexlab{}.
\newblock \emph{\bibinfo{title}{The synthesizer programming problem: improving the usability of sound synthesizers}}.
\newblock \bibinfo{thesistype}{Master's\ thesis}. \bibinfo{school}{University of Victoria}.
\newblock


\bibitem[Tawalbeh et~al\mbox{.}(2020)]%
        {tawalbeh2020iot}
\bibfield{author}{\bibinfo{person}{Lo’ai Tawalbeh}, \bibinfo{person}{Fadi Muheidat}, \bibinfo{person}{Mais Tawalbeh}, {and} \bibinfo{person}{Muhannad Quwaider}.} \bibinfo{year}{2020}\natexlab{}.
\newblock \showarticletitle{IoT Privacy and security: Challenges and solutions}.
\newblock \bibinfo{journal}{\emph{Applied Sciences}} \bibinfo{volume}{10}, \bibinfo{number}{12} (\bibinfo{year}{2020}), \bibinfo{pages}{4102}.
\newblock


\bibitem[Team et~al\mbox{.}(2024)]%
        {team2024gemma}
\bibfield{author}{\bibinfo{person}{Gemma Team}, \bibinfo{person}{Thomas Mesnard}, \bibinfo{person}{Cassidy Hardin}, \bibinfo{person}{Robert Dadashi}, \bibinfo{person}{Surya Bhupatiraju}, \bibinfo{person}{Shreya Pathak}, \bibinfo{person}{Laurent Sifre}, \bibinfo{person}{Morgane Rivi{\`e}re}, \bibinfo{person}{Mihir~Sanjay Kale}, \bibinfo{person}{Juliette Love}, {et~al\mbox{.}}} \bibinfo{year}{2024}\natexlab{}.
\newblock \showarticletitle{Gemma: Open models based on gemini research and technology}.
\newblock \bibinfo{journal}{\emph{arXiv preprint arXiv:2403.08295}} (\bibinfo{year}{2024}).
\newblock


\bibitem[Touvron et~al\mbox{.}(2023)]%
        {touvron2023llama}
\bibfield{author}{\bibinfo{person}{Hugo Touvron}, \bibinfo{person}{Louis Martin}, \bibinfo{person}{Kevin Stone}, \bibinfo{person}{Peter Albert}, \bibinfo{person}{Amjad Almahairi}, \bibinfo{person}{Yasmine Babaei}, \bibinfo{person}{Nikolay Bashlykov}, \bibinfo{person}{Soumya Batra}, \bibinfo{person}{Prajjwal Bhargava}, \bibinfo{person}{Shruti Bhosale}, {et~al\mbox{.}}} \bibinfo{year}{2023}\natexlab{}.
\newblock \showarticletitle{Llama 2: Open foundation and fine-tuned chat models}.
\newblock \bibinfo{journal}{\emph{arXiv preprint arXiv:2307.09288}} (\bibinfo{year}{2023}).
\newblock


\bibitem[Tsiakoulis et~al\mbox{.}(2012)]%
        {tsiakoulis2012statistical}
\bibfield{author}{\bibinfo{person}{Pirros Tsiakoulis}, \bibinfo{person}{Milica Ga{\v{s}}ic}, \bibinfo{person}{Matthew Henderson}, \bibinfo{person}{Joaquin Planells-Lerma}, \bibinfo{person}{Jorge Prombonas}, \bibinfo{person}{Blaise Thomson}, \bibinfo{person}{Kai Yu}, \bibinfo{person}{Steve Young}, {and} \bibinfo{person}{Eli Tzirkel}.} \bibinfo{year}{2012}\natexlab{}.
\newblock \showarticletitle{Statistical methods for building robust spoken dialogue systems in an automobile}.
\newblock \bibinfo{journal}{\emph{Proceedings of the 4th applied human factors and ergonomics}} (\bibinfo{year}{2012}).
\newblock


\bibitem[Upadhyay et~al\mbox{.}(2023)]%
        {upadhyay2023studying}
\bibfield{author}{\bibinfo{person}{Pooja Upadhyay}, \bibinfo{person}{Sharon Heung}, \bibinfo{person}{Shiri Azenkot}, {and} \bibinfo{person}{Robin~N Brewer}.} \bibinfo{year}{2023}\natexlab{}.
\newblock \showarticletitle{Studying exploration \& long-term use of voice assistants by older adults}. In \bibinfo{booktitle}{\emph{Proceedings of the 2023 CHI Conference on Human Factors in Computing Systems}}. \bibinfo{pages}{1--11}.
\newblock


\bibitem[Vaswani et~al\mbox{.}(2017)]%
        {vaswani2017attention}
\bibfield{author}{\bibinfo{person}{Ashish Vaswani}, \bibinfo{person}{Noam Shazeer}, \bibinfo{person}{Niki Parmar}, \bibinfo{person}{Jakob Uszkoreit}, \bibinfo{person}{Llion Jones}, \bibinfo{person}{Aidan~N Gomez}, \bibinfo{person}{{\L}ukasz Kaiser}, {and} \bibinfo{person}{Illia Polosukhin}.} \bibinfo{year}{2017}\natexlab{}.
\newblock \showarticletitle{Attention is all you need}.
\newblock \bibinfo{journal}{\emph{Advances in neural information processing systems}}  \bibinfo{volume}{30} (\bibinfo{year}{2017}).
\newblock


\bibitem[Wang et~al\mbox{.}(2022)]%
        {wang2022self}
\bibfield{author}{\bibinfo{person}{Yizhong Wang}, \bibinfo{person}{Yeganeh Kordi}, \bibinfo{person}{Swaroop Mishra}, \bibinfo{person}{Alisa Liu}, \bibinfo{person}{Noah~A Smith}, \bibinfo{person}{Daniel Khashabi}, {and} \bibinfo{person}{Hannaneh Hajishirzi}.} \bibinfo{year}{2022}\natexlab{}.
\newblock \showarticletitle{Self-instruct: Aligning language models with self-generated instructions}.
\newblock \bibinfo{journal}{\emph{arXiv preprint arXiv:2212.10560}} (\bibinfo{year}{2022}).
\newblock


\bibitem[Weld et~al\mbox{.}(2022)]%
        {weld2022survey}
\bibfield{author}{\bibinfo{person}{Henry Weld}, \bibinfo{person}{Xiaoqi Huang}, \bibinfo{person}{Siqu Long}, \bibinfo{person}{Josiah Poon}, {and} \bibinfo{person}{Soyeon~Caren Han}.} \bibinfo{year}{2022}\natexlab{}.
\newblock \showarticletitle{A survey of joint intent detection and slot filling models in natural language understanding}.
\newblock \bibinfo{journal}{\emph{Comput. Surveys}} \bibinfo{volume}{55}, \bibinfo{number}{8} (\bibinfo{year}{2022}), \bibinfo{pages}{1--38}.
\newblock


\bibitem[Xu et~al\mbox{.}(2022)]%
        {xu2022systematic}
\bibfield{author}{\bibinfo{person}{Frank~F Xu}, \bibinfo{person}{Uri Alon}, \bibinfo{person}{Graham Neubig}, {and} \bibinfo{person}{Vincent~Josua Hellendoorn}.} \bibinfo{year}{2022}\natexlab{}.
\newblock \showarticletitle{A systematic evaluation of large language models of code}. In \bibinfo{booktitle}{\emph{Proceedings of the 6th ACM SIGPLAN International Symposium on Machine Programming}}. \bibinfo{pages}{1--10}.
\newblock


\bibitem[Yang and Newman(2013)]%
        {yang2013learning}
\bibfield{author}{\bibinfo{person}{Rayoung Yang} {and} \bibinfo{person}{Mark~W Newman}.} \bibinfo{year}{2013}\natexlab{}.
\newblock \showarticletitle{Learning from a learning thermostat: lessons for intelligent systems for the home}. In \bibinfo{booktitle}{\emph{Proceedings of the 2013 ACM international joint conference on Pervasive and ubiquitous computing}}. \bibinfo{pages}{93--102}.
\newblock


\bibitem[Yates et~al\mbox{.}(2003)]%
        {yates2003reliable}
\bibfield{author}{\bibinfo{person}{Alexander Yates}, \bibinfo{person}{Oren Etzioni}, {and} \bibinfo{person}{Daniel Weld}.} \bibinfo{year}{2003}\natexlab{}.
\newblock \showarticletitle{A reliable natural language interface to household appliances}. In \bibinfo{booktitle}{\emph{Proceedings of the 8th international conference on Intelligent user interfaces}}. \bibinfo{pages}{189--196}.
\newblock


\end{thebibliography}

%\appendix

\end{document}